\documentclass[aps,pra,twocolumn,showpacs]{revtex4-1}
\usepackage{graphicx,amsmath,amssymb}
\usepackage[usenames]{color}

\begin{document}

\title{
Symmetry breaking patterns, tricriticalities and quadruple points in quantum Rabi model with bias and nonlinear interaction
}

\author{Zu-Jian Ying }
\email{yingzj@lzu.edu.cn}
%\email{zujianying@yahoo.com}
%\email{zying@unisa.it}
%\email{zjying@csrc.ac.cn}
\affiliation{School of Physical Science and Technology, Lanzhou 730000, China}
%\affiliation{CNR-SPIN and Dipartimento di Fisica ``E. R. Caianiello'', Universit\`a di Salerno, 84084 Fisciano, Italy}

\begin{abstract}
Quantum Rabi model (QRM) is fascinating not only because of its broad relevance and but also due to its few-body quantum phase transition. In practice both the bias and the nonlinear coupling in QRM are important controlling parameters in experimental setups. We study the interplay of the bias and the nonlinear interaction with the linear coupling in the ground state which exhibits various patterns of symmetry breaking and different orders of transitions. Several situations of tricriticality are unveiled in the low frequency limit and at finite frequencies. We find that the full quantum-mechanical effect leads to novel transitions, tricriticalities and quadruple points, which are much beyond the semiclassical picture. We clarify the underlying mechanisms by analyzing the energy competitions and the essential changeovers of the quantum states, which enables us to extract most analytic phase boundaries.
\end{abstract}
\pacs{ }
\maketitle

%\date{\today}

%\begin{widetext}
%\end{widetext}

\section{Introduction}

In the past decade, both experimental \cite{Diaz2019RevModPhy} and
theoretical \cite{Braak2011,Boite2020} progresses have brought the strong
light-matter interaction to the frontiers of quantum optics and quantum
physics. The experimental access to increasingly larger coupling strengths
has opened a new regime with\ a rich phenomenology \cite%
{Diaz2019RevModPhy,Kockum2019NRP} unexpected in weak couplings. Beyond the
Jaynes-Cummings model \cite{JC-model} which is valid in weak couplings, the
quantum Rabi model (QRM) \cite{rabi1936} is the most fundamental model for
strong light-matter interaction. The QRM also has a wide relevance, being a
fundamental building block for quantum information and quantum computation
\cite{Diaz2019RevModPhy,Romero2012}, closely connected to models in condense
matter \cite{Kockum2019NRP}, and even applied in black hole physics \cite%
{Pedernales-PRL-2018}. Theoretically, the milestone work of revealing Braak
integrability \cite{Braak2011} for the QRM\ has not only heated up the
interest in the light-matter interaction but also triggered an intense
dialogue between mathematics and physics \cite%
{Boite2020,Wolf2012,Solano2011,exact_chenqh,Irish2014,PRX-Xie-Anistropy,Batchelor2015,Hwang2015PRL, Ying2015,LiuM2017PRL,CongLei2017,CongLei2019,Ying-2018-arxiv,Liu2015,Liu2017JPA, Ashhab2013,ChenGang2012,FengMang2013,ZhangYY2016,Yimin2018,AnistropicShen2017,Villamizar-2019, Cong-2020,XieQ-2017JPA,Zhong-2013JPA,Eckle-2017JPA,Penna-2018JPA,Maciejewski-2019JPA,XieYF-2019JPA}.

The fast experimental advances have pushed the coupling strength all the way
through from weak-, strong-coupling regimes to ulstrastrong-coupling regime
and even beyond\cite%
{Diaz2019RevModPhy,Wallraff2004,Gunter2009,Niemczyk2010,Peropadre2010,FornDiaz2017, Forn-Diaz2010,Scalari2012,Xiang2013,Yoshihara2017NatPhys,Yoshihara2017,Kockum2017}%
. A most fascinating consequence of continuing enhancements of the coupling
strength is the emergence of phase-transition-like phenomena \cite%
{Ashhab2013,Ashhab2010,Hwang2015PRL,
Ying2015,LiuM2017PRL,CongLei2017,Ying-2018-arxiv,Liu2015,Liu2017JPA}. As a
usual impression, phase transitions mostly occur in thermodynamical limit in
condensed matter. Note that the QRM is composed of a single qubit or
spin-half system in coupling with a light field or a bosonic mode, thus the
few-body quantum phase transition found in the QRM appears quite particular.
Interestingly via the scaling relation of the critical behavior it has been
established that the few-body phase transition can be can be bridged to the
phase transition in the thermodynamic limit \cite{LiuM2017PRL}.

Along with the continuing regime expanding of the QRM in the frontiers of
quantum optics and quantum physics, a playground for novel physics in
nonlinear quantum optics is also opened by an extended version of the QRM,
so-called two-photon quantum Rabi model
\cite{Felicetti2018-mixed-TPP-SPP,Simone2018,Felicetti2015-TwoPhotonProcess,Puebla2017-TwoPhotonProcess,Travenec2012,
e-collpase-Lo-1998,e-collpase-Duan-2016,e-collpase-Garbe-2017,CongLei2019}.
The conventional QRM is a linear model in the sense that it is via a
single-photon process of absorption and emission for the qubit or spin-half
system to couple with a bosonic mode. The interaction in the two-photon
model involves a coupling via two-photon process of absorption and emission,
which is nonlinear. Recently the nonlinear two-photon interaction has
attracted an increasing attention as the model can be implemented in
trapped-ion systems \cite%
{Felicetti2015-TwoPhotonProcess,Puebla2017-TwoPhotonProcess} and
superconducting circuits \cite{Felicetti2018-mixed-TPP-SPP,Simone2018} with
the interaction strength enhanced to realize the ultrastrong regime.
Critical behavior also appears in such two-photon QRM and a special
phenomenon is the spectral collapse \cite%
{Felicetti2015-TwoPhotonProcess,e-collpase-Lo-1998,e-collpase-Duan-2016,e-collpase-Garbe-2017},
i.e., its discrete spectrum collapses into a continuous band when the
nonlinear interaction strength approaches to the critical point. It has been noticed that the
spectral collapse can be tuned from incomplete collapse to complete collapse
by variation of the system frequency \cite{CongLei2019}.

An important character of the QRM noteworthy to mention is the symmetry. It
is well-known that the QRM has the so-called parity symmetry. Generally
speaking, it is quite common that only at certain parameter point can a
physical system possess a symmetry and one needs very fine-tuned conditions
to maintain the symmetry, while the realistic conditions in experimental
setups may break the symmetry. Nevertheless, although symmetry is the
diamond of physics, what makes the world of physics really rich is often the
symmetry breaking. As far as the QRM in the light-matter interaction is
concerned, it is known that anisotropy \cite{LiuM2017PRL,PRX-Xie-Anistropy}
in the coupling will preserve the parity symmetry. However the existence of
a bias or a nonlinear interaction will definitely break the parity symmetry
of the linear QRM. Despite that a pure two-photon model also has a parity
symmetry, the mixture of the single-photon coupling and two-photon
interaction will break both the parity symmetries of the linear QRM and of
the two-photon QRM. In such a mixed case novel phenomena could arise, such
as the emergence of triple point and spontaneous symmetry breaking\cite%
{Ying-2018-arxiv}. Recently there is a trend of growing interest in the
mixed model \cite%
{Ying-2018-arxiv,Casanova2018npj,Xie2019,PengJie2019,FelicettiPRL2020}.
So far, most the studies have been focusing on the mixed model without
taking the bias into account, however in realistic conditions of
experimental setups it is more general to have both the bias and nonlinear
interaction in the presence \cite{Bertet-Nonlinear-Experim-Model-2005}. In
such a situation a full knowledge of the competition and interplay of the
bias and nonlinear interaction is still lacking and very desirable.

In this work we present a systematic study on a general realistic model \cite%
{Felicetti2018-mixed-TPP-SPP,Bertet-Nonlinear-Experim-PRL-2005,Bertet-Nonlinear-Experim-Model-2005}
comprised of the linear coupling, the bias, nonlinear interaction as well as
a nonlinear Stark term. We focus on the ground state which exhibits various
patterns of symmetry breaking and different orders of phase transitions. We
find that in such a realistic model tricritical-like behavior can be induced
in diverse situations. It is also interesting to get a contrast of
semiclassical picture in the low frequency limit and the full
quantum-mechanical effect at finite frequencies. We demonstrate that the
quantum-mechanical effect leads to a much richer phenomenology including
novel transitions, tricriticalities and quadruple points.

The paper is organized as follows. In Section \ref{Sect-model} we introduce
the general model with bias and nonlinear interaction. The parity symmetry
in the conventional QRM is addressed. In Section \ref{Sect-Breaking-Patterns}
we show different patterns of symmetry breaking that the model exhibits. In
Section \ref{Sect-PhaseDiagram-low-w} \ we present the full phase diagrams
in low frequency limit, in the respective or simultaneous presence of the
bias and the nonlinear interaction, together with obtained analytic
boundaries. \ In the low frequency limit we reveal a first tricriticality in
Section \ref{Sect-TriCri-low-w-limit}. In Section \ref{SectionAll-Finte-w}
we discuss the finite-frequency case, unveiling four more situations of
tricriticalities. We show that there could be three, even four successive
transitions, the analytic phase boundaries are also presented. Quadruple
points are demonstrated in Section \ref{Sect-quadruple}. In Section \ref%
{Sect-WaveFunction} the essential changes of the wave function is
illustrated for the phase transitions. Section \ref{Sect-Mechanisms} is
devoted to clarify the mechanisms underlying the various symmetry breaking
patterns, different orders of phase transitions, successive transitions and
different responses of physical quantities to the transitions. We address
the semiclassical picture and the full quantum mechanics effect, the latter
leading to more phase transitions and thus being the origin of the various
tricriticalities and quadruple points. We also the explain the scaling of
the Stark term in the nonlinear interaction. Section \ref{Sect-Analytic}
provides brief derivations of the analytic boundaries. In the final section
we summarize the results and discuss the realization regime for experimental
parameters in superconducting circuit systems.

\section{Model and parity symmetry}

\label{Sect-model}

Besides the linear coupling of the QRM, experimental setups in
superconducting circuits actually involve both nonlinear coupling and bias,
with a Hamiltonian reading as \cite%
{Felicetti2018-mixed-TPP-SPP,Bertet-Nonlinear-Experim-Model-2005}
\begin{eqnarray}
H &=&H_{0}+H_{t}+H_{\epsilon }  \nonumber \\
H_{0} &=&\omega a^{\dag }a+\frac{\Omega }{2}\sigma _{x}+g_{1}\sigma _{z}{%
(a^{\dag }+a)} \\
H_{t} &=&g_{2}\sigma _{z}{\left[ (a^{\dag })^{2}+a^{2}+\chi \widetilde{n}%
\right] ,\quad }H_{\epsilon }=-\epsilon \sigma _{z}  \nonumber
\end{eqnarray}%
where $\sigma _{x,y,z}$ is the Pauli matrix, $a^{\dagger }(a)$ creates
(annihilates) a bosonic mode with frequency $\omega .$ The $\Omega $ term is
atomic level splitting in cavity\ systems, while in the superconducting
circuit systems it is tunneling between the spin-up and spin-down states of
the flux qubit \cite{flux-qubit-Mooij-1999} as represented by $\sigma _{z}$.
Following Ref.\cite{Irish2014}, we adopt the spin notation in the
superconducting circuit systems which can realize very strong couplings. The
conventional QRM is described by $H_{0}$ where the coupling is linear, via
the single-photon process of absorption and emission, with a coupling
strength $g_{1}$. The nonlinear interaction is denoted by $H_{t}$ with the
coupling strength $g_{2}$. Here we have included a Stark-like term \cite%
{Felicetti2018-mixed-TPP-SPP,Eckle-2017JPA}, $\chi \widetilde{n}$ with\ $%
\widetilde{n}=a^{\dag }a+aa^{\dag }$ essentially being the photon number, to
retrieve the conventional two-photon form \cite{Felicetti2018-mixed-TPP-SPP}
by $\chi =0$ and\ the quadratic form ${(a^{\dag }+a)}^{2}$ in experimental
setups \cite{Bertet-Nonlinear-Experim-Model-2005} by $\chi =1$. One can also
obtain a pure Stark-like term\cite{Eckle-2017JPA} by setting $%
g_{2}\rightarrow 0$ while keeping\ $\chi $ inversely proportional to the
bare nonlinear interaction $\chi \propto 1/g_{2}$. It turns out that for the
properties discussed in present work the Stark-like term contributes to a
scaling factor and by%
\begin{equation}
\widetilde{g}_{2}=\left( 1+\chi \right) g_{2}  \label{scaled-g2}
\end{equation}%
we get similar results. For simplicity, unless otherwise mentioned, we use $%
g_{2}$ to represent general $\widetilde{g}_{2}$ throughout the figures. The
origin of the scaling will be clarified in Section \ref{Sect-Scaling-Stark}.

The conventional QRM $H_{0}$ possesses the parity symmetry $\hat{P}=-\sigma
_{x}(-1)^{a^{\dagger }a}$ which commutes with $H_{0}$. The parity operation $%
\hat{P}$ simultaneously reverses the spin sign and inverses the effective
spatial space $x\rightarrow -x$. The spin sign reversion can be seen
directly as $\sigma _{x}=\left( \sigma ^{+}+\sigma ^{-}\right) .$ The space
inversion can be conveniently shown by expanding the wave function on the
basis of quantum harmonic oscillator $\left\vert n\right\rangle $, $%
\left\vert \Psi \right\rangle =\left\vert \Psi _{+}\right\rangle +\left\vert
\Psi _{-}\right\rangle =\sum_{n=0}^{\infty }\left( c_{n,+}\left\vert
n,+\right\rangle +c_{n,-}\left\vert n,-\right\rangle \right) $, where $+$ ($%
- $) labels the up (down) spin in $z$ direction. Then the action of the
parity operation leads to $\hat{P}\left\vert \Psi \right\rangle =$ $%
\sum_{n=0}^{\infty }\left( -1\right) ^{n}\left( c_{n,+}\left\vert
n,-\right\rangle +c_{n,-}\left\vert n,+\right\rangle \right) $. In the
spatial coordinate it means the transform
\begin{eqnarray*}
\hat{P} &:&\Psi _{\pm }(x)\rightarrow \sum_{n}\left( -1\right) ^{n}c_{n,\mp
}\phi _{n}\left( x\right) \\
\quad &=&\sum_{n}c_{n,\mp }\phi _{n}\left( -x\right) =\Psi _{\mp }(-x),
\end{eqnarray*}%
where we have applied the fact that the eigenstate of quantum harmonic
oscillator, $\phi _{n}\left( x\right) $, is an odd (even) function of $x$
for an odd (even) quantum number $n$. Thus we see the space inversion $%
x\rightarrow -x$, besides the spin reversion. The parity symmetry requires\ $%
\Psi ^{(\pm )}(x)=P\Psi ^{(\pm )}(-x)$ where $P=\pm 1$. The ground state of
QRM has a parity $P=-1$. Apparently, either in the negative or positive
parity symmetry, the spin expectation $\left\langle \sigma _{z}\right\rangle
$ along $z$ direction is vanishing, which is a characteristic of the parity
symmetry. In the present work we focus on the symmetry breaking in the
ground state of the general model with the bias, the nonlinear interaction
and the Stark coupling.

\section{Different patterns of symmetry breaking}
\label{Sect-Breaking-Patterns}

%%%%%%%%%%%%%%%%%%%%%%%%%%%%%%%%%%%%%%%%%%%%%%%%%%%%%%%%%%%%%%%%%%%%%%%%%%%%%%%%%%%%%%%%%%%%%%%%%%
\begin{figure}[tbp]
\includegraphics[width=1.0\columnwidth]{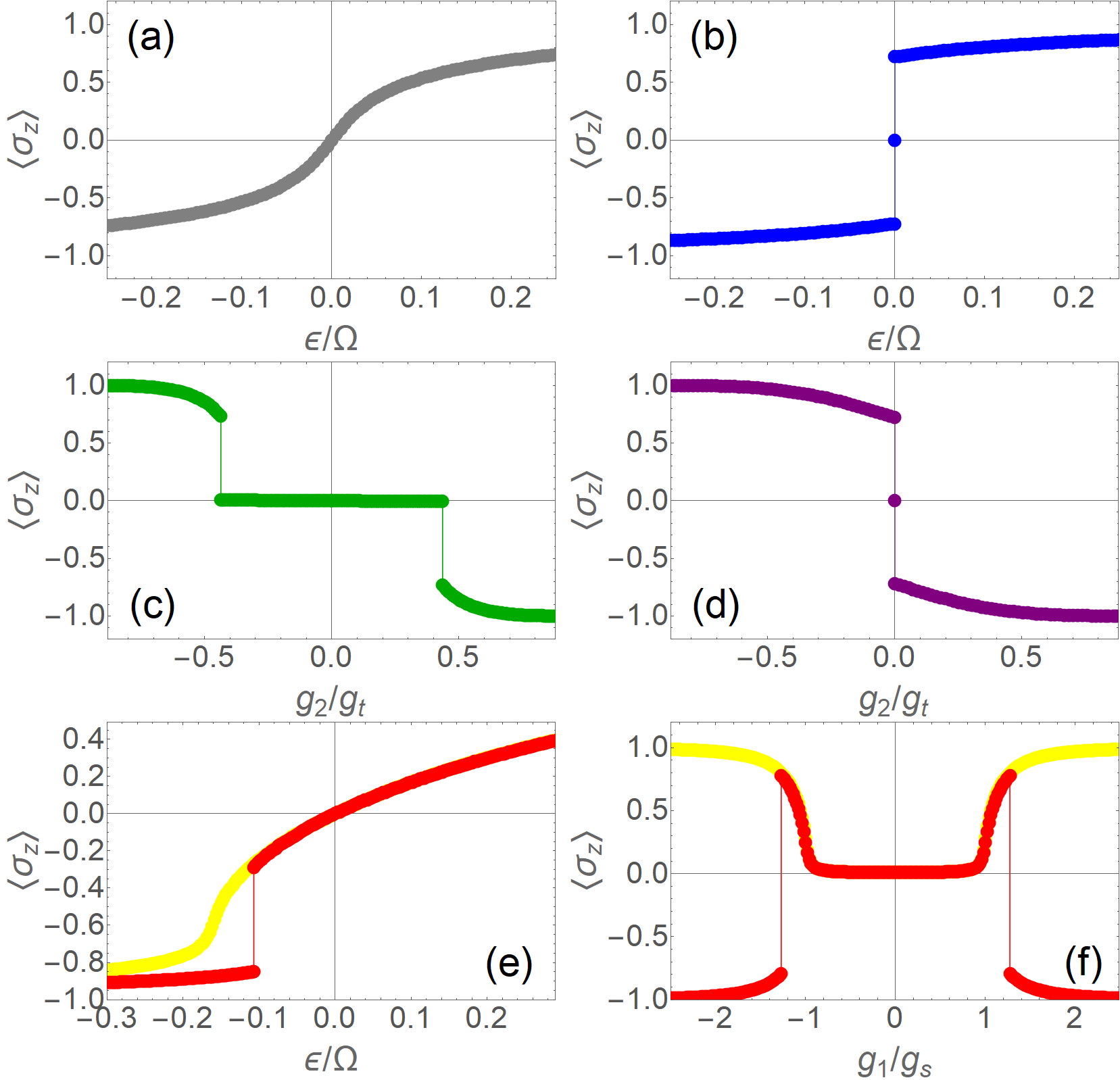}
\caption{(color online) \textit{Different patterns of symmetry breaking.}
(a-f) Spin expectation $\langle \protect\sigma _{z}\rangle $ in the cases of
(a) paramagnetic-like, (c) antiferromagnetic-like, (b,d) the spontaneous
symmetry breaking, (e) paramagnetic+first/second-order-transition and (f)
antiferromagnetic+first/second-order-transition. These cases are illustrated
by fixed parameters at $g_{2}=0$ with $g_{1}=0.9g_{\mathrm{s}}$ (a) or $%
g_{2}=0$ (b), at $\protect\epsilon =0$ with $g_{1}=0.9g_{\mathrm{s}}$ (c) or
$g_{1}=1.2g_{\mathrm{s}}$ (d), at $g_{1}=0.7g_{\mathrm{s}}$ (e) with $%
g_{2}=0.6g_{\mathrm{t}}$ (green) or $g_{2}=0.65g_{\mathrm{t}}$ (red), at $%
\protect\epsilon =1g_{\mathrm{t}}$ (f) with $g_{2}=-0.02g_{\mathrm{t}}$
(green) or $g_{2}=0.02g_{\mathrm{t}}$ (red), given a frequency $\protect%
\omega =0.001\Omega $. }
\label{Fig-patterns}
\end{figure}
%%%%%%%%%%%%%%%%%%%%%%%%%%%%%%%%%%%%%%%%%%%%%%%%%%%%%%%%%%%%%%%%%%%%%%%%%%%%%%%%%%%%%%%%%%%%%%%%%%

%%%%%%%%%%%%%%%%%%%%%%%%%%%%%%%%%%%%%%%%%%%%%%%%%%%%%%%%%%%%%%%%%%%%%%%%%%%%%%%%%%%%%%%%%%%%%%%%%%
\begin{figure}[tbp]
\includegraphics[width=1.0\columnwidth]{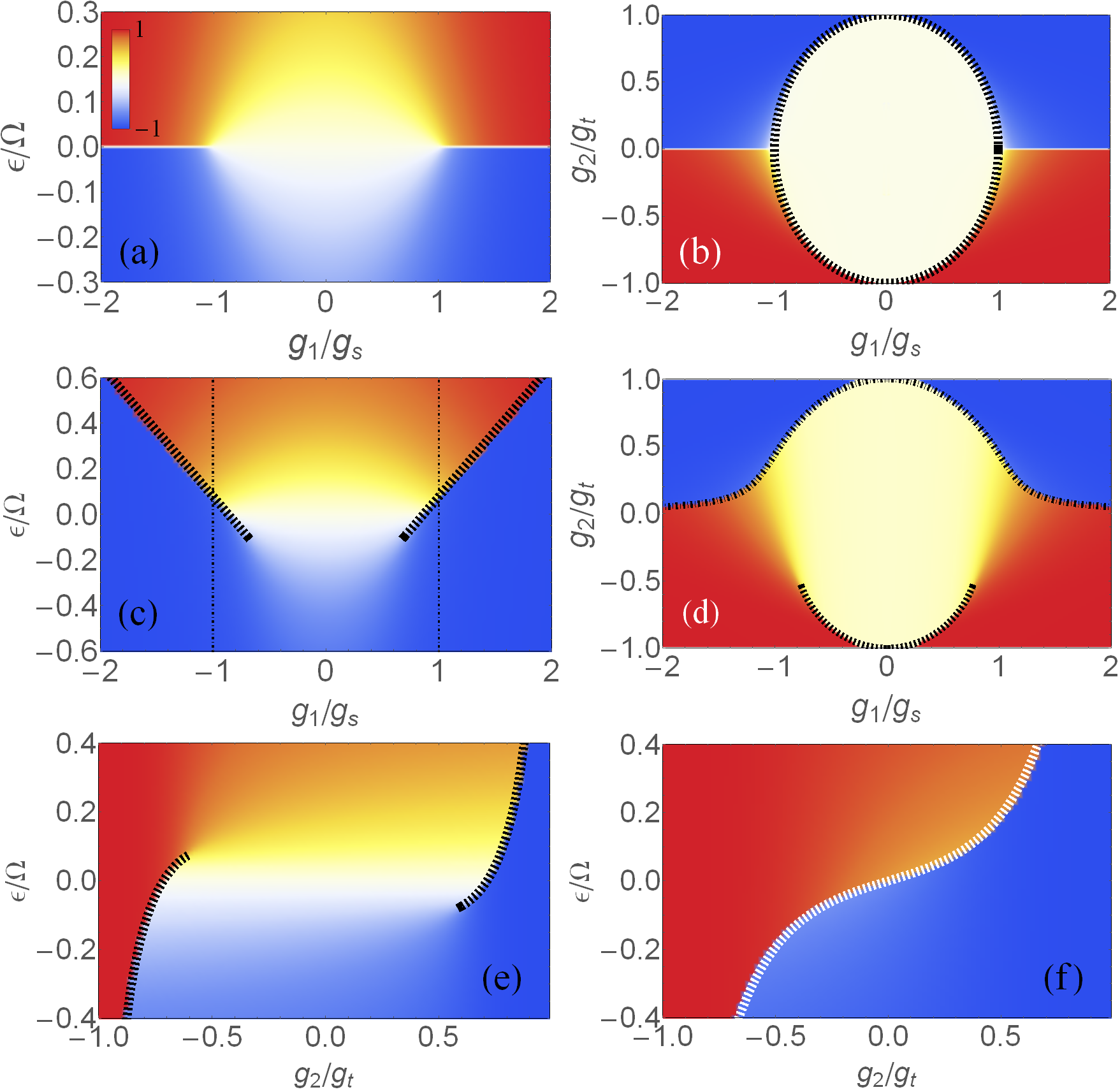}
\caption{(color online) \textit{Phase diagrams and analytic phase boundaries
in low frequency limit.} Spin expectation $\langle \protect\sigma_z\rangle$
at a fixed parameter (a) $g_2=0$, (b) $\protect\epsilon =0$, (c) $g_2=0.5g_%
\mathrm{t}$, (d) $\protect\epsilon=10g_\mathrm{t}$, (e) $g_1=0.7g_\mathrm{s}$%
, (f) $g_1=1.2g_\mathrm{s}$. Here $\protect\omega =0.01\Omega$. All panels
share the same color legend for $\langle \protect\sigma_z\rangle$ as (a).
The dashed or dot-dashed curves are analytic boundaries
\eqref{g1cSemiclassical} and \eqref{hzcSemiclassical}, the vertical lines in
(c) are marking $g_\mathrm{s}$ as a reference for the boundary moving.}
\label{Fig-SemiClassicDiagrams}
\end{figure}
%%%%%%%%%%%%%%%%%%%%%%%%%%%%%%%%%%%%%%%%%%%%%%%%%%%%%%%%%%%%%%%%%%%%%%%%%%%%%%%%%%%%%%%%%%%%%%%%%%

Either the bias and the nonlinear interaction will break the parity symmetry
of the linear QRM $H_{0}$. Interestingly different scenarios arise in the
interplay of linear coupling with the bias and the nonlinear interaction,
leading to various patterns of symmetry breaking. On the one hand, the
linear QRM has a critical point at $g_{1}=g_{\mathrm{s}}\equiv \sqrt{\omega
\Omega }/2$ which also turns out to be a critical point for change of
symmetry-breaking patterns (though, by a perspective view in Section \ref%
{Sect-PhaseDiagram-low-w}, this pattern critical point may be shifted when
both the bias and the nonlinear interaction are present). The regimes below
and above the critical point respond to the symmetry breaking with
completely different sensitivities. On the other hand, within a same
linear-coupling regime, the processes of symmetry breaking may be
essentially different in the presence of the bias and nonlinear interaction.

Fig.\ref{Fig-patterns} illustrates the different patterns of symmetry
breaking in response to the bias and the nonlinear interaction, as
calculated from exact diagonalization. Fig.\ref{Fig-patterns} (a) shows the
evolution of the spin expectation $\left\langle \sigma _{z}\right\rangle $
with respect to the variation of the bias, below $g_{\mathrm{s}}$ of linear
coupling and in the absence of the nonlinear interaction. We see that the
amplitude of $\left\langle \sigma _{z}\right\rangle $ increases gradually
with the bias strength, which is paramagnetic-like. Fig.\ref{Fig-patterns}
(c) shows the dependence of $\left\langle \sigma _{z}\right\rangle $ on the
strength of the nonlinear interaction below $g_{\mathrm{s}}$. We see that $%
\left\langle \sigma _{z}\right\rangle $ has no response to the nonlinear
interaction $g_{2}$ and remains vanishing until the strength of $g_{2}$
reaches some critical point $g_{2c}$. Once $g_{2}$ goes beyond $g_{2c},$ the
spin expectation $\left\langle \sigma _{z}\right\rangle $ jumps abruptly to
a finite value and then starts approaching to saturation. This pattern with
a threshold for polarization is antiferromagnetic-like. We see the
essentially different patterns of symmetry breaking: there is a first-order
phase transition induced by the nonlinear interaction, while there is no
transition in introducing the bias. Above the critical point $g_{\mathrm{s}}$
of the linear coupling, both the bias and nonlinear interaction bring
another pattern. In Fig.\ref{Fig-patterns} (b,d), we see that a tiny
strength of either the bias or the nonlinear coupling will lead to dramatic
change in $\left\langle \sigma _{z}\right\rangle $ which jumps to a finite
value. This is the pattern of spontaneous symmetry breaking. This means the
system is extremely sensitive to the perturbation of the bias or the
nonlinear interaction, in a sharp contrast to both the paramagnetic-like
pattern and antiferromagnetic-like pattern. Furthermore, as in Fig.\ref%
{Fig-patterns} (e), the interplay of the bias and the nonlinear interaction
may lead to a paramagnetic-like pattern followed by a second-order-like
transition (green line) or first-order transition (red line) . It is also
interesting to see that in the interplay with both the bias and nonlinear
interaction increasing $g_{1}$ could bring about an antiferromagnetic-like
pattern but with the afore-mentioned first-order transition replaced by a
second-order transition, as illustrated by the green line in Fig.\ref%
{Fig-patterns} (f). This occurs for the opposite signs of the bias and the
nonlinear interaction. When the signs are the same another pattern could
emerge, i.e. antiferromagnetic-like pattern plus successive transitions of
second-order and first-order kinds, as shown by the red line in Fig.\ref%
{Fig-patterns} (f).

\section{Phase diagrams and analytic transition boundaries in the low
frequency limit}
\label{Sect-PhaseDiagram-low-w}

To get a perspective view we plot the phase diagrams in the full parameter
spaces, as in Fig. \ref{Fig-SemiClassicDiagrams}. Panels (a) shows the
dependence of $\left\langle \sigma _{z}\right\rangle $ on the bias and the
linear coupling, in the absence of the nonlinear interaction. The spin
expectation $\left\langle \sigma _{z}\right\rangle $ has a positive value in
the red region for $\epsilon >0$ and a negative value in the blue region for
$\epsilon <0$. The white line at $\epsilon =0,$ with vanishing $\left\langle
\sigma _{z}\right\rangle ,$ is parity-symmetry line from the conventional
QRM. As we see, for the weak linear coupling regime $\left\vert
g_{1}\right\vert <\left\vert g_{\mathrm{s}}\right\vert ,$ when the bias is
getting stronger, the color gradually turns from white to red or blue, which
indicates no phase transition. Differently, for the whole strong coupling
regime $\left\vert g_{1}\right\vert >\left\vert g_{\mathrm{s}}\right\vert $,
there is a sharp color change across the parity-symmetry line, indicating
the spontaneous symmetry breaking. Panel (b) shows the behavior of $%
\left\langle \sigma _{z}\right\rangle $ in the interplay of the nonlinear
interaction and the linear coupling in the absence of the bias. We see that,
besides the parity-symmetry white line at $g_{2}=0$, another white round
region is opened where the parity symmetry for the ground state is also
unbroken. The antiferromagnetic-like pattern occurs in the regime of the
round region. The dashed line along the circumference of the round region in
panel (b) is the analytic boundary
\begin{equation}
\left\vert g_{1c}\right\vert =g_{\mathrm{s}}\sqrt{1-\widetilde{g}_{2}^{2}/g_{%
\mathrm{t}}^{2}}  \label{g1C-pure-g2}
\end{equation}%
where $\widetilde{g}_{2}=\left( 1+\chi \right) g_{2}$, which reproduces the
numerical boundary.

Fig. \ref{Fig-SemiClassicDiagrams} (c,d) illustrate the mutual influence of
the bias and the nonlinear interaction over their phase diagrams. Panel (c)
is plotted in the dimensions of the bias and the linear coupling, in the
presence of a finite nonlinear interaction $g_{2}=0.7g_{\mathrm{t}}.$ Here $%
g_{\mathrm{t}}=\omega /2$ is the physical limit for the nonlinear
interaction, beyond the limit the system energy becomes negatively unbound
thus being unphysical. We see that, in the presence of a finite nonlinear
interaction, the boundary of $\left\langle \sigma _{z}\right\rangle $ gets
tilted from the horizontal line in zero-$g_{2}$ case. Furthermore, the
transition boundary enters the regime $[-g_{\mathrm{s}},g_{\mathrm{s}}]$
where originally there is no transition in the absence of the nonlinear
interaction. Panel (d) is plotted in the $g_{1}$-$g_{2}$ section in the
presence of a finite strength of the bias $\epsilon =0.1\Omega .$ We see for
$g_{2}>0$ the connection of the circle and the horizontal line originally in
$\epsilon =0$ case (panel (b)) now becomes round, with the boundary changing
from a dome shape to be a hill shape. In this reshaping the transition at
the boundary remains to be first-order. For $g_{2}<0,$ some section of the
first-order round boundary disappears, with the jump of $\left\langle \sigma
_{z}\right\rangle $ closed and softened, turning the original half-circle
boundary to be an arc shape. Let us label by $g_{2}^{\mathrm{E}}$ the
critical nonlinear interaction for the ends of the arc boundary$.$ The
analytic expression of $g_{2}^{\mathrm{E}}$ will be given in Eq.\eqref{g2E}
and the dependence on the bias strength plotted in Fig.\ref{Fig-Saddle-flat}%
(b) in Section \ref{Sect-Analytic-low-w}. Meanwhile, the arc spanning angle
gets narrower than the half circle, i.e., $\left\vert g_{1c}\right\vert $ is
smaller at the same value of $g_{2}$. The rest first-order arc boundary,
remaining in the large-$g_{2}$-amplitude regime, shrinks with an enhanced
bias.

Fig. \ref{Fig-SemiClassicDiagrams} (e,f) show the phase diagrams at a fixed
linear coupling below $g_{\mathrm{s}}$ (panel (e)) or above $g_{\mathrm{s}}$
(panel (f)). Below $g_{\mathrm{s}}$ there are two first-order boundaries in
the variations of the bias and the nonlinear interaction, which are
separated. When the linear coupling get stronger the two boundaries are
curved, with their ends getting closer and finally connected to form one
first-order boundary above $g_{\mathrm{s}}$.

For a quantitative description, we extract the analytic boundary marked by $%
g_{1c}$ or $\epsilon _{c}$ as follows
\begin{eqnarray}
\left\vert g_{1c}\right\vert &=&g_{\mathrm{s}}[1+\frac{g_{\mathrm{t}%
}\epsilon }{\widetilde{g}_{2}\Omega }]\sqrt{1-\widetilde{g}_{2}^{2}/g_{%
\mathrm{t}}^{2}},  \label{g1cSemiclassical} \\
\epsilon _{c} &=&\frac{\widetilde{g}_{2}}{g_{\mathrm{t}}}[\frac{\left\vert
g_{1}\right\vert /g_{\mathrm{s}}}{\sqrt{1-\widetilde{g}_{2}^{2}/g_{\mathrm{t}%
}^{2}}}-1]\ \Omega .  \label{hzcSemiclassical}
\end{eqnarray}%
We leave the analytic derivation in Section \ref{Sect-Analytic-low-w}. Note
here, whereas for $g_{\mathrm{t}}\epsilon >0$ the extension of boundary is
unlimited, for $g_{\mathrm{t}}\epsilon >0$ the validity regime is $%
\left\vert g_{2}\right\vert <g_{2}^{\mathrm{E}}$ which is the arc boundary.
We leave the detail of derivations in Section \ref{Sect-Analytic}. Setting $%
\epsilon =0$ retrieves the round boundary $\left\vert g_{1c}\right\vert =g_{%
\mathrm{s}}\sqrt{1-\widetilde{g}_{2}^{2}/g_{\mathrm{t}}^{2}}$ in the absence
of bias \cite{Ying-2018-arxiv}. We plot the analytic boundaries by the
dashed or dot-dashed lines in Fig. \ref{Fig-SemiClassicDiagrams} (b-f),
which are in good agreements with the numerical boundaries.

\section{Tricriticality-(i) in the low frequency limit}
\label{Sect-TriCri-low-w-limit}

%%%%%%%%%%%%%%%%%%%%%%%%%%%%%%%%%%%%%%%%%%%%%%%%%%%%%%%%%%%%%%%%%%%%%%%%%%%%%%%%%%%%%%%%%%%%%%%%%%
\begin{figure}[tbp]
\includegraphics[width=1.0\columnwidth]{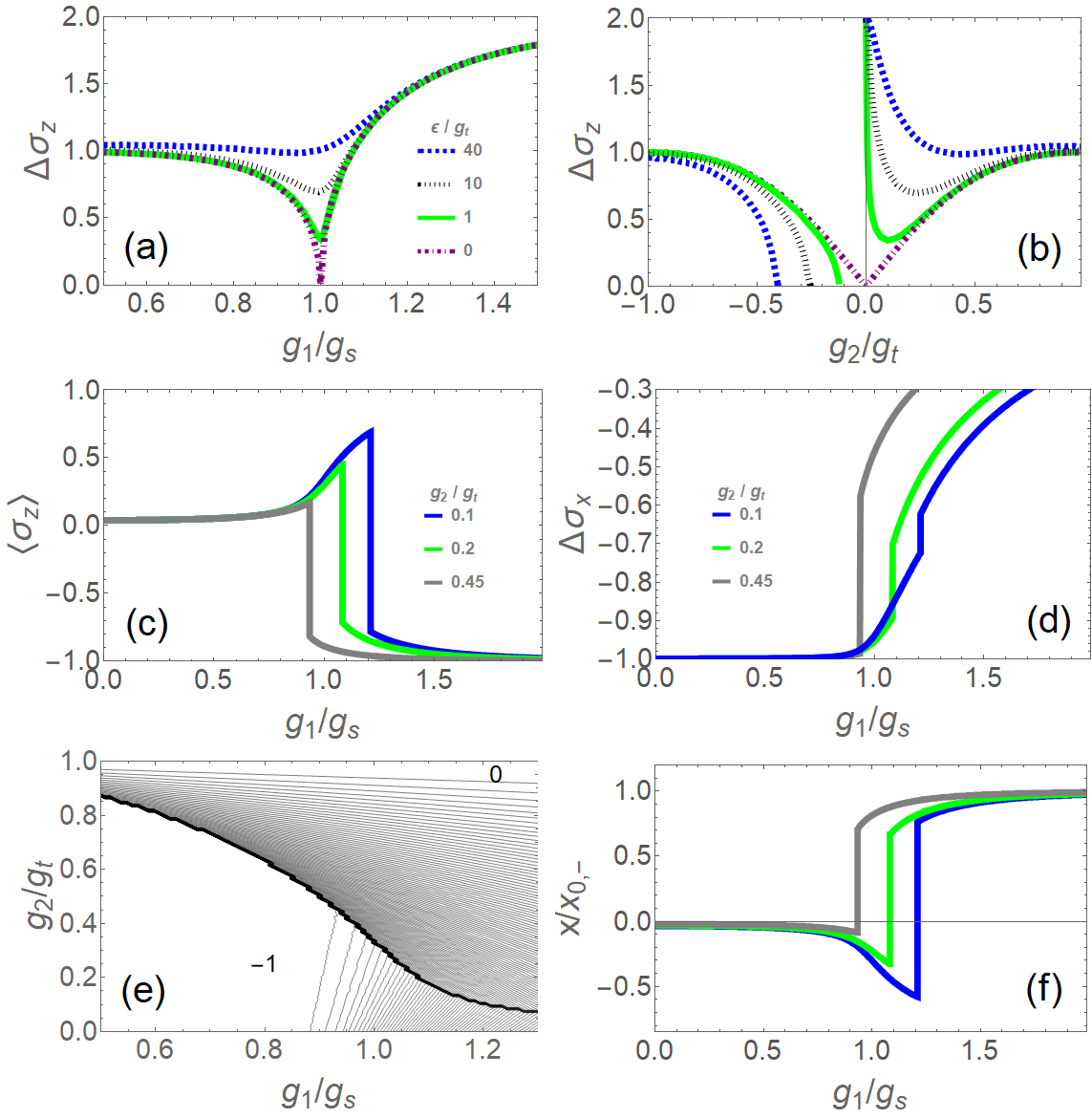}
\caption{(color online) \textit{Tricriticality-(i) in low frequency limit.}
(a) Spin expectation discontinuity $\Delta\protect\sigma_z$ along the
transition boundary in $g_2>0$ regime as a function of $g_1$ for $\protect%
\epsilon= 0g_\mathrm{t}$ (purple dot-dashed), $1g_\mathrm{t}$ (green solid),
$10g_\mathrm{t}$ (black dotted) and $40g_\mathrm{t}$ (blue dashed) at $%
\protect\omega=0.001\Omega$. (b) $\Delta\protect\sigma_z$ as a function of $%
g_2$. (c-f) Three-phase behavior of $\langle \protect\sigma_z\rangle$ (c), $%
\langle \protect\sigma_x\rangle$ (d,e), $\langle \widehat{x} \rangle$ (f)
for $\protect\epsilon= 40g_\mathrm{t}$. (c,d,f) give illustrations with
fixed $g_2=0.1g_\mathrm{t}$ (blue), $g_2=0.2g_\mathrm{t}$ (green) and $%
g_2=0.45g_\mathrm{t}$ (gray), while (e) is a contour plot. }
\label{Fig-Spin-jump}
\end{figure}
%%%%%%%%%%%%%%%%%%%%%%%%%%%%%%%%%%%%%%%%%%%%%%%%%%%%%%%%%%%%%%%%%%%%%%%%%%%%%%%%%%%%%%%%%%%%%%%%%%

In Fig. \ref{Fig-SemiClassicDiagrams}(b) one may notice on each side $%
g_{1}=\pm g_{\mathrm{s}}$ is a tricritical point where the round boundary
and horizontal line are crossing. In Fig. \ref{Fig-Spin-jump}(a) by the
purple dot-dashed line we show the spin expectation discontinuity $\Delta
\sigma _{z}$, i.e. the jump of $\langle \sigma _{z}\rangle $ across the
boundary, with the finite value of $\Delta \sigma _{z}$ representing the
first-order transition. The transition becomes second-order at $g_{\mathrm{s}%
}$ as indicated by the vanishing of $\Delta \sigma _{z}$. \ In the presence
of the bias, this second-order transition also turns to be first-order, as
we illustrate by $\epsilon /\Omega =0.001$, $0.01$ and $0.04$ in the low
frequency limit ($\epsilon /g_{\mathrm{t}}=1,$ $10$ and $40$ if taking $%
\omega =0.001\Omega $)$.$ With the bias increasing, the shape of the $\Delta
\sigma _{z}$ minimum evolves from a sharp dip into a round valley. Fig. \ref%
{Fig-Spin-jump}(b) provides a view of $\Delta \sigma _{z}$ in the $g_{2}$
dimension, which includes both boundaries in the positive and negative $%
g_{2} $ regimes. The $\langle \sigma _{z}\rangle $-vanishing point at zero
bias is extending into a window at finite biases. In the negative $g_{2}$
regime the remaining finite-$\langle \sigma _{z}\rangle $ section in panel
(b) corresponds to the boundary arc. \ In the positive $g_{2}$ regime, it is
worthwhile to follow the evolution of the minimum position $\Delta \sigma
_{z}$, which is moving away from the original point $g_{2}=0$. As
afore-mentioned, this minimum point is originally a tricritical point in the
absence of the bias, now in the presence of the bias it turns out to be an
imprint of new tricriticality.

Indeed, when scanning $g_{1}$ in Fig. \ref{Fig-Spin-jump}(c), as
demonstrated by the case $g_{2}=0.1g_{\mathrm{t}}$ (blue line) we see a
three-phase-like scenario: firstly a flat region in spin expectation $%
\langle \sigma _{z}\rangle $, secondly a fast-rising region, finally jumping
into a region with opposite sign. The three phases look more distinct in the
evolution of the spin expectation in $x$ direction, $\langle \sigma
_{x}\rangle ,$ as shown by the blue line in Fig. \ref{Fig-Spin-jump}(d).
Essential changes of the three phases may be indicated by $\langle
a^{+}+a\rangle $ which is effective spatial particle position $x$ (we shall
discuss more in Sections \ref{Sect-WaveFunction} and \ref{Sect-Mechanisms}).
As shown in Fig. \ref{Fig-Spin-jump}(f), the\ effective particle resides
closely around the origin in the first phase, moves obviously away from
origin in the second phase and jumps abruptly to the other side in the third
phase. These three phases are separated by two transition-like points, the
first transition is second-order-like and the second one is of first order.
When the bias strength increases, the two transitions get closer to each
other and finally meet, as illustrated by $g_{2}=0.2g_{\mathrm{t}}$ (green
lines) and $g_{2}=0.45g_{\mathrm{t}}$ (gray lines) in Fig. \ref%
{Fig-Spin-jump}(c,d,f). Such a scenario of two separate transitions
converging to one transition forms a tricritical-like point, which can be
seen more clearly by the contour plot of $\langle \sigma _{x}\rangle $ in
Fig. \ref{Fig-Spin-jump}(e). This tricritical-like point is located around
the afore-mentioned $\Delta \sigma _{z}$ minimum position.

\section{Novel tricriticalities at finite frequencies }
\label{SectionAll-Finte-w}

The low frequency limit discussed in previous sections is also the
semiclassical limit, as the wave-packet size is so small that it can be
regarded as a semiclassical mass point (see Section \ref{Sect-Semiclassical}%
). \ In such a semiclassical limit, \ in each quadrant of the phase diagram
the nonlinear interaction induces only one transition in the absence of the
bias and at most two transitions in the interplay with the bias. On the
other hand, the spontaneous symmetry breaking occurs immediately upon any
tiny strength of the bias or nonlinear interaction. In this section, we
shall see that the full quantum-mechanical effect\ at finite frequencies
will change this picture and lead to richer scenarios. We find that
additional transitions appear, novel tricriticalities arise and the
spontaneous symmetry breaking exhibits a fine structure.

\subsection{Additional transition and Tricriticality-(ii) induced by the
frequency respectively in the bias or the non-linear interaction}
\label{Sect-finite-w}

%%%%%%%%%%%%%%%%%%%%%%%%%%%%%%%%%%%%%%%%%%%%%%%%%%%%%%%%%%%%%%%%%%%%%%%%%%%%%%%%%%%%%%%%%%%%%%%%%%
\begin{figure}[tbp]
\includegraphics[width=1.0\columnwidth]{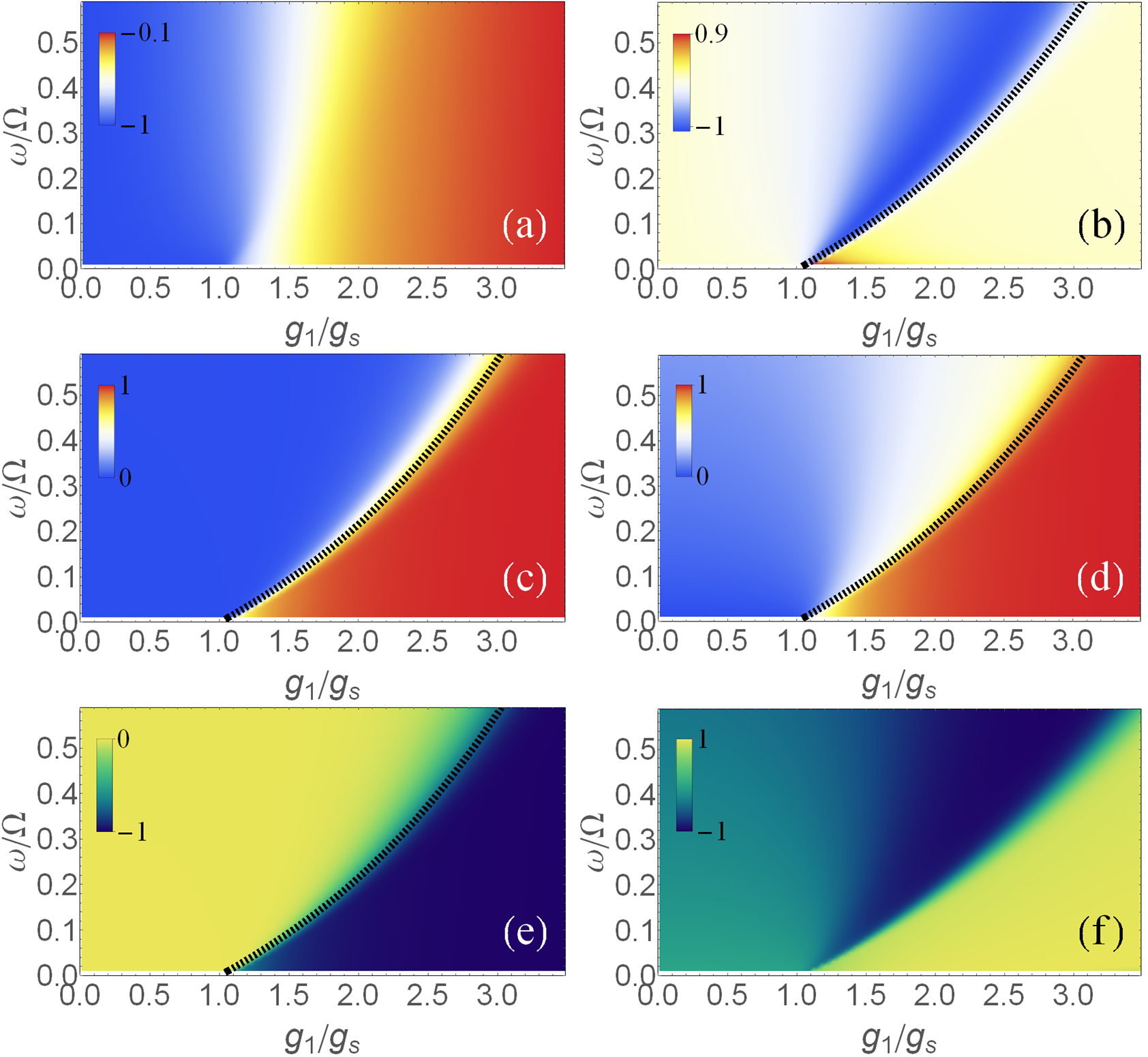}
\caption{(color online) \textit{Tricriticality-(ii) induced by frequency $%
\protect\omega$.} Phase diagrams of different quantities by frequency
variation ($\protect\omega$) at $\protect\epsilon=0.001g_\mathrm{t}$ with $%
g_2=0$ (a,c,e) and $g_2=0.0001g_\mathrm{t}$ with $\protect\epsilon=0$
(b,d,f) for (a) $\langle \protect\sigma_x\rangle$, (b) $\langle \hat{x}
\rangle_+$, (c) $\langle \protect\sigma_z\rangle$, (d) $\langle \hat{x}
\rangle_- /|x_{0,-}|$ ($\hat{x}=(a+a^\dagger )/\protect\sqrt{2}$), (e) $%
\langle \hat{x} \rangle/|x_{0,+}|$, (f) $\tilde{x}_+ =\langle \hat{x}
\rangle _+ /(|x_{0,+}|\protect\rho_+)$. The dashed lines in (b-e) are $g^{%
\mathrm{II}}_{1c}$ extracted from analytic Eqs.\eqref{hzcII-0} and
\eqref{g2cII-0}.}
\label{Fig-Frequency}
\end{figure}
%%%%%%%%%%%%%%%%%%%%%%%%%%%%%%%%%%%%%%%%%%%%%%%%%%%%%%%%%%%%%%%%%%%%%%%%%%%%%%%%%%%%%%%%%%%%%%%%%%

The tricriticality in Section \ref{Sect-TriCri-low-w-limit} occurs in the
low frequency limit. The transitions and the tricriticality arise from the
competition and interplay of the bias, the nonlinear interaction and the
linear coupling. In such a situation, different physical quantities exhibit
imprints of each transition at a same transition point, as one can see from
Fig. \ref{Fig-Spin-jump}(c,d,f). Here we show another kind of tricriticality
induced by the frequency which has a different nature and transition
positions diverge for different physical quantities.

In Fig.\ref{Fig-Frequency}, we show a variety of physical quantities with
the dependence on the frequency $\omega $, under a fixed bias $\epsilon
=0.1\Omega $ in panels (a,c,e) or a fixed nonlinear interaction $%
g_{2}=0.01g_{\mathrm{t}}$ in panels (b,d,f). The two parameter cases have
similar behavior despite some detail and sign difference for some
quantities. As expected, in the low frequency limit both $\langle \sigma
_{x}\rangle $ (panel (a)) and $\langle \sigma _{z}\rangle $ (panel (c)) show
one transition at a same point around $g_{1}=g_{\mathrm{s}}$. However, when
the frequency is raised, $\langle \sigma _{x}\rangle $ and $\langle \sigma
_{z}\rangle $ respond differently. In fact, the transition in $\langle
\sigma _{x}\rangle $ is not much affected by the frequency except for some
softening of the transition, whereas the transition in $\langle \sigma
_{z}\rangle $ is moving obviously toward the larger-$g_{1}$ direction. The
diverging evolutions of the transition positions of $\langle \sigma
_{x}\rangle $ and $\langle \sigma _{z}\rangle $ indicate an additional
transition induced by the finite frequency, thus the one transition in the
low-frequency limit becomes two transitions at finite frequencies. It also
seems peculiar that the spin expectations $\langle \sigma _{x}\rangle $ and $%
\langle \sigma _{z}\rangle $ respond to the two transitions respectively:
the first transition induces response in $\langle \sigma _{x}\rangle $ but
leaves no imprints in $\langle \sigma _{z}\rangle $, while the second
transition releases a strong onset signal in $\langle \sigma _{z}\rangle $
but gives no sign in $\langle \sigma _{x}\rangle .$ This additional
transition can also be seen from the effective particle position or
displacement $\langle \widehat{{x}}{\rangle =}\langle {a^{\dag }+a\rangle /}%
\sqrt{2}$, as shown in Fig.\ref{Fig-Frequency}(c).

The two diverging transitions can be also detected simultaneously by a
single physical quantity, such as the spin-filtered displacement $\langle
\widehat{{x}}{\rangle }_{\pm }{=}\langle {a^{\dag }+a\rangle }_{\pm }{/}%
\sqrt{2}$ which only counts the contribution from one spin component, as
shown in \ref{Fig-Frequency}(b,d)\ where it is quite clear to visualize two
boundaries corresponding to the two transitions. To have unified upper and
lower bounds for plotting we also introduce the normalized spin-filtered
displacement $\widetilde{x}_{\pm }=\langle {a^{\dag }+a\rangle }_{\pm }/(%
\sqrt{2}\rho _{\pm }\left\vert x_{0,sign(-\widetilde{g}_{2})}\right\vert )$,
where $\rho _{\pm }=\langle \psi ^{\pm }|\psi ^{\pm }\rangle =\left( 1\pm
\left\langle \sigma _{z}\right\rangle \right) /2$ is the spin-component
weight and $x_{0,\pm }=\mp g_{1}^{\prime }/(1\pm \widetilde{g}_{2}^{\prime
}) $ is the potential displacement (see Section \ref{Sect-Mechanisms}). Here
we have defined $g_{1}^{\prime }=\sqrt{2}g_{1}/\omega \ $\ and $\widetilde{g}%
_{2}^{\prime }=2\widetilde{g}_{2}/\omega $. Besides the normalization, $%
\widetilde{x}_{\pm }$ have another convenience that it has three regimes of
values respectively for the three phases separated by the two transitions.
Thus the three phases can can be distinguished by three colors, as shown in
Fig.\ref{Fig-Frequency}(f). It should be mentioned\ that at higher
frequencies there is some discrepancy for the second transition point from$\
\widetilde{x}_{\pm }$. This spurious transition discrepancy is simply coming
from the cancellation effect around the transition from its numerator $%
\langle {a^{\dag }+a\rangle }_{\pm }$ and denominator $\rho _{\pm }$, while
separately both $\langle {a^{\dag }+a\rangle }_{\pm }$ and $\rho _{\pm }$
have the right second transition point. \ Nevertheless, the discrepancy at
low frequencies is negligible so we can still use it for further discussions
by the advantages of its normalization and value(color)-phase correspondence.

Reversely in lowering the frequency, the two boundaries of the three phases
will converge to one point, thus forming a new triple point and another kind of tricriticality.
Conventionally a tricriticality is composed of three critical boundaries,
while this tricriticality here is comprised of two critical boundaries at finite
frequencies and one critical point at zero frequency. This critical point connects
two phases which does not adjoin directly through either of the other two boundaries.
Thus there are three kinds of critical behavior. In this sense we still term it as a tricriticality.
It should be noted that the tricriticality of this case, as labeled by (ii),
is distinguished from Tricriticality-(i) in Section \ref%
{Sect-TriCri-low-w-limit}. Tricriticality-(i) happens in the presence of
both the bias and the nonlinear interaction, while Tricriticality-(ii) here
occurs in the respective presence of the bias or the nonlinear interaction.
From the mechanism clarification in Section \ref{Sect-Mechanisms} we will
see that the additional transition and new tricriticality originate from a
full-quantum-mechanical effect, in a contrast to the semiclassical effect in
the low frequency limit.

\subsection{Tricriticality-(iii) induced by the bias or the nonlinear
interaction at finite frequencies}

%%%%%%%%%%%%%%%%%%%%%%%%%%%%%%%%%%%%%%%%%%%%%%%%%%%%%%%%%%%%%%%%%%%%%%%%%%%%%%%%%%%%%%%%%%%%%%%%%%
\begin{figure}[tbp]
\includegraphics[width=1.0\columnwidth]{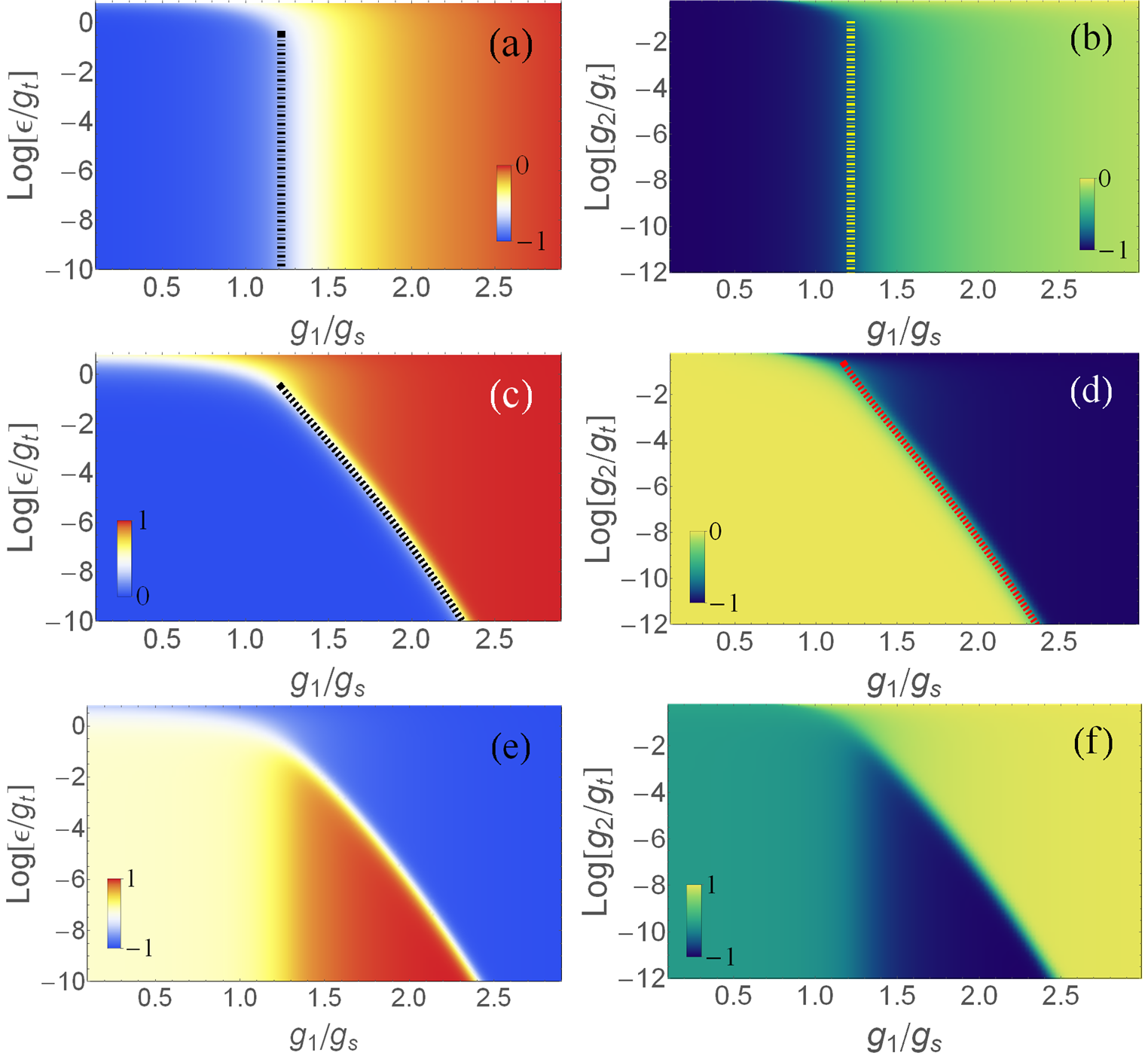}
\caption{(color online) \textit{Tricriticality-(iii) induced by the bias or
the nonlinear interaction at a finite frequency.} Phase diagrams at $\protect%
\omega =0.1\Omega $ for variation of $\protect\epsilon $ at $g_{2}=0$: (a) $%
\langle \protect\sigma _{x}\rangle $, (c) $\langle \protect\sigma %
_{z}\rangle $, (e) $\widetilde{x}_{-}$. Phase diagrams for variation of $%
g_{2}$ at $\protect\epsilon =0$: (b) $\langle \protect\sigma _{x}\rangle $,
(d) $\langle \protect\sigma _{z}\rangle $, (f) $\widetilde{x}_{+}$. The
dot-dashed lines in (a,b) are analytic $g_{1c}^{\mathrm{I}}$ and the dashed
lines in (c,d) are analytic $\protect\epsilon _{c}^{\mathrm{II}}$ and $%
g_{2c}^{\mathrm{II}}$ in Eqs. \eqref{hzcII-0} and \eqref{g2cII-0}}
\label{Fig-tricritical-1}
\end{figure}
%%%%%%%%%%%%%%%%%%%%%%%%%%%%%%%%%%%%%%%%%%%%%%%%%%%%%%%%%%%%%%%%%%%%%%%%%%%%%%%%%%%%%%%%%%%%%%%%%%

%%%%%%%%%%%%%%%%%%%%%%%%%%%%%%%%%%%%%%%%%%%%%%%%%%%%%%%%%%%%%%%%%%%%%%%%%%%%%%%%%%%%%%%%%%%%%%%%%%
\begin{figure}[tbp]
\includegraphics[width=1.0\columnwidth]{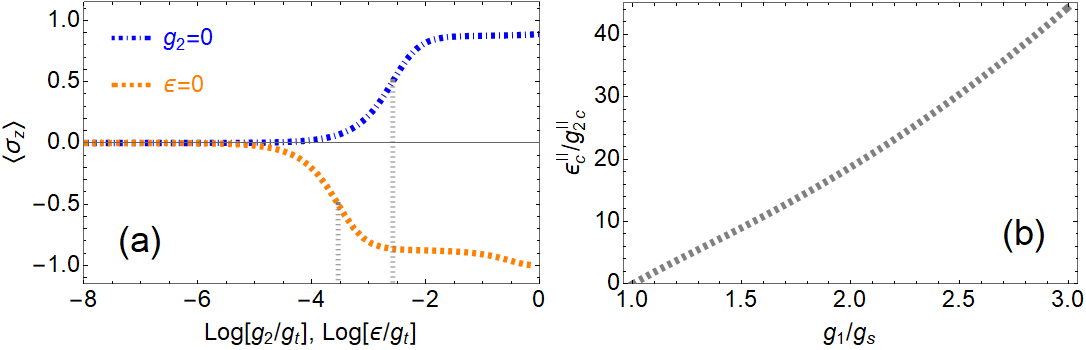}
\caption{(color online) \textit{Sensitivity competition for spontaneous
symmetry breaking.} (a) $\langle \protect\sigma_{z}\rangle $ depending on $%
\protect\epsilon$ at $g_2=0$ (orange dashed) and on $g_2 $ at $\protect%
\epsilon=0$ (blue dot-dashed) at $g_1=1.5g_\mathrm{s}$ and $\protect\omega %
=0.1\Omega$. (b) Threshold ratio $\protect\epsilon _{c}^{\mathrm{II}} /
g_{2c}^{\mathrm{II}} $ depending on $g_1$}
\label{Fig-sensativity}
\end{figure}
%%%%%%%%%%%%%%%%%%%%%%%%%%%%%%%%%%%%%%%%%%%%%%%%%%%%%%%%%%%%%%%%%%%%%%%%%%%%%%%%%%%%%%%%%%%%%%%%%%

It will provide another view by fixing a finite frequency and varying the
bias or the nonlinear interaction. As described in Section \ref%
{Sect-PhaseDiagram-low-w}, in the low frequency limit we see from Fig. \ref%
{Fig-SemiClassicDiagrams} (a,b) that in each quadrant of the phase diagrams
there is no more than one transition$.$ As revealed in Section \ref%
{Sect-finite-w}, at a finite frequency the single transition turns to be two
successive transitions. The variation of the bias or the nonlinear
interaction will influence the transitions and induce a third tricriticality
which we label by Tricriticality-(iii).

The successive transitions can be seen more clearly from a zoom-in view by
a logarithm scale for the variations of $\epsilon $ and $g_{2},$ as
illustrated by Fig.\ref{Fig-tricritical-1} at a finite frequency $\omega
=0.1\Omega $. Panels (a,c,e) present the phase diagrams for the pure bias
dependence without the nonlinear interaction and panels (b,d,f) for the
nonlinear interaction in the absence of the bias. To distinguish the two
transitions we label the transition in $\langle \sigma _{x}\rangle $ by $%
g_{1c}^{\mathrm{I}}$ and that in $\langle \sigma _{z}\rangle $ by $g_{1c}^{%
\mathrm{II}}.$ We see in panels (a,b) that the first transition in $\langle
\sigma _{x}\rangle $ does not vary at weak strengths of $\epsilon $ or $%
g_{2} $, except that the transition point at low frequencies shifts a bit
from $g_{1c}\sim g_{\mathrm{s}}$ to $g_{1c}^{\mathrm{I}}\approx \sqrt{\omega
^{2}+\sqrt{\omega ^{4}+g_{\mathrm{s}}^{4}}}$\cite{Ying2015} due to the width
of wave packet in the wave-packet splitting. \ In a sharp contrast, the
second transition is very sensitive to the variation of the bias and the
nonlinear interaction. In fact, as demonstrated by Fig.\ref%
{Fig-tricritical-1}(c,d), the transition point $g_{1c}^{\mathrm{II}}$ has an
exponential dependence on $\epsilon $ and $\Omega $. Analytically we find
the second boundary as a function of $g_{1}$ (see the derivation in Section %
\ref{Sect-Analytic}):
\begin{eqnarray}
\left\vert \epsilon _{c}^{\mathrm{II}}\right\vert &=&\frac{(1-t)\Omega }{%
4\delta _{c}\zeta }\exp [-\frac{\zeta ^{2}\overline{g}_{1}^{2}\Omega }{%
2\omega }],\quad \text{for }g_{2}=0,  \label{hzcII-0} \\
\left\vert \widetilde{g}_{2c}^{\mathrm{II}}\right\vert &=&\frac{(1-t)g_{%
\mathrm{t}}}{\delta _{c}\zeta ^{3}\overline{g}_{1}^{2}}\exp [-\frac{\zeta
^{2}\overline{g}_{1}^{2}\Omega }{2\omega }],\quad \text{for }\epsilon =0,
\label{g2cII-0}
\end{eqnarray}%
where $\ \overline{g}_{1}\equiv g_{1}/g_{\mathrm{s}}$, $\delta _{c}=e^{-1},$
$t=(1-\zeta )^{2}/2+\omega /(\overline{g}_{1}^{2}\Omega )$ and $\zeta =(1-%
\overline{g}_{1}^{-4})^{1/2}$. The analytic boundaries $\epsilon _{2c}^{%
\mathrm{II}}$ and $g_{2c}^{\mathrm{II}}$\ are plotted as the dashed lines in
Fig.\ref{Fig-tricritical-1}(c,d), in good agreements with the numerical
results.

With the strength increase of the bias or the nonlinear interaction, the two
transitions respectively reflected in $\langle \sigma _{x}\rangle $ and $%
\langle \sigma _{z}\rangle $ \ are getting closer and finally meet to form
Tricriticality-(iii). A better view of this tricriticality can be obtained
from $\widetilde{x}_{\pm }$ as in Fig.\ref{Fig-tricritical-1}(e,f) where
three phases are distinctly represented by three colors. Above the
tricritical point it is one transition of first-order type, while below the
tricritical point the transition is bifurcated into a second-order-like
transition and a first-order-like one. Exactly speaking, in the bias case
the transition above the tricritical point is a short extension from the
first-order transition below the tricritical point. This transition soon
gets softened and fades away when the linear coupling $g_{1}$ is reduced to
below $g_{\mathrm{s}}$. In the nonlinear interaction case, the
first-order-like transition covers the entire $g_{2}$ regime thus also the
whole $g_{1}$ regime.

To distinguish from Tricritcalities (i) and (ii) let us mention the
difference. Tricritcality-(i) in the low frequency limit revealed in Section %
\ref{Sect-TriCri-low-w-limit} occurs in the presence of both the bias and
the nonlinear interaction. Tricriticality-(iii) here needs only the bias or
the nonlinear interaction. Tricriticality-(ii) unveiled in Section \ref%
{Sect-finite-w} is induced by the variation of the frequency, here
Tricriticality-(iii) is induced by the bias or nonlinear interaction at a
fixed finite frequency.

%\begin{widetext}
%%%%%%%%%%%%%%%%%%%%%%%%%%%%%%%%%%%%%%%%%%%%%%%%%%%%%%%%%%%%%%%%%%%%%%%%%%%%%%%%%%%%%%%%%%%%%%%%%%
\begin{figure*}[tbph]
\centering
\includegraphics[width=0.99\textwidth,scale=1.00]{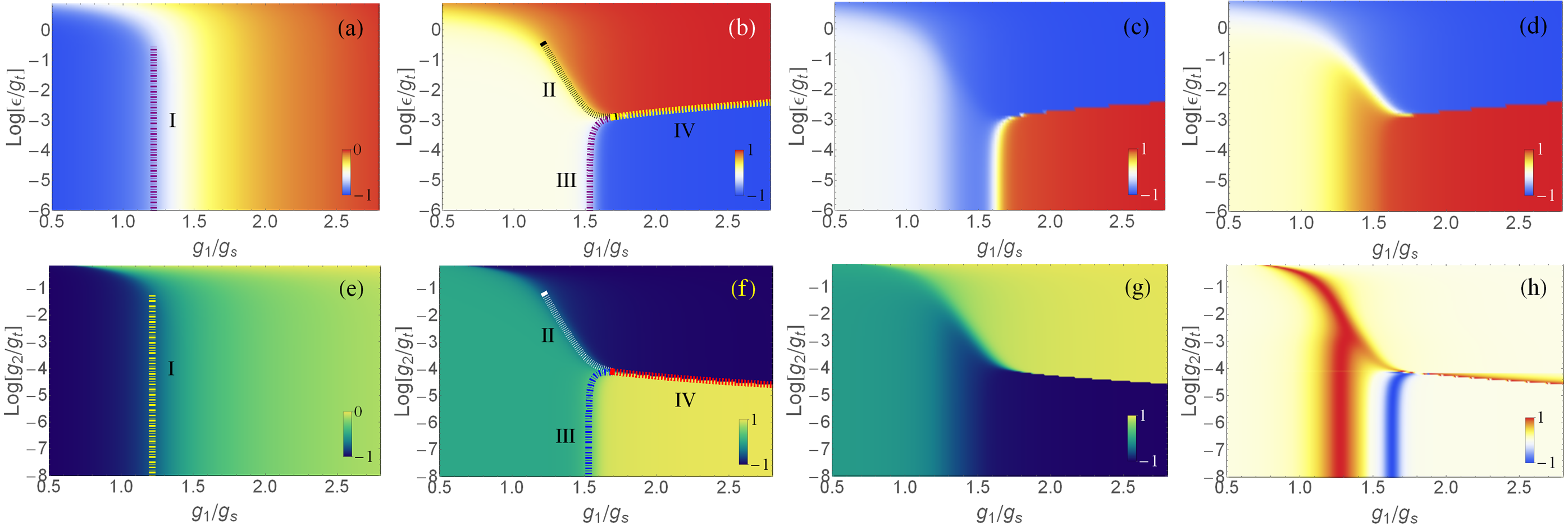}
\caption{(color online) \textit{Tricriticality-(iv) with two
tricritical-like points in the interplay of the bias the nonlinear
interaction.} Phase diagrams at $\protect\omega =0.1\Omega $ for variation
of $\protect\epsilon $ at $g_{2}=10^{-4}g_{t}$: (a) $\langle \protect\sigma %
_{x}\rangle $, (b) $\langle \protect\sigma _{z}\rangle $, (c) $\widetilde{x}%
_{+}$, (c) $\widetilde{x}_{-}$. Density plot for variation of $g_{2}$ at $%
\protect\epsilon =10^{-3}g_{t}$: (e) $\langle \protect\sigma _{x}\rangle $,
(f) $\langle \protect\sigma _{z}\rangle $, (g) $\widetilde{x}_{+}$, (h) $d%
\widetilde{x}_{-}/dg_{1}$ scaled by the local peak amplitude. The lines are
our analytic $g_{1c}^{\mathrm{I}}$ in panels (a,e), $\protect\epsilon _{c}^{%
\mathrm{II}}$, $\protect\epsilon _{c}^{\mathrm{III}}$, $\protect\epsilon %
_{c}^{\mathrm{IV}}$ in panel (b) and $g_{2c}^{\mathrm{II}}$ $g_{2c}^{\mathrm{%
III}}$, $g_{2c}^{\mathrm{IV}}$ in panel (f) (see Eqs. \eqref{hzC-II}-
\eqref{g2cIV}).}
\label{Fig-tricritical-2}
\end{figure*}
%%%%%%%%%%%%%%%%%%%%%%%%%%%%%%%%%%%%%%%%%%%%%%%%%%%%%%%%%%%%%%%%%%%%%%%%%%%%%%%%%%%%%%%%%%%%%%%%%%
%\end{widetext}

The scenario of Tricriticality-(iii) \ also gives rise to a fine structure
of the spontaneous symmetry breaking for the finite frequency case. Note
that the negative-$\epsilon $/$g_{2}$ regime has the same tricritical
scenario, except for being antisymmetric for $\langle \sigma _{z}\rangle $
and symmetric for $\langle \sigma _{z}\rangle $ in the quadrants of the
phase diagrams. Thus, rather than an immediate jump of $\langle \sigma
_{x}\rangle $ from zero to a finite value upon the opening of the bias or
the nonlinear interaction, there is now a window within which the parity
symmetry of the ground state is maintained to some large extent, as
indicated by the vanishing $\langle \sigma _{z}\rangle $. Out of the window
the symmetry is broken. This window becomes narrower when the linear
coupling gets stronger, but can be widened by a higher frequency.

\subsection{Sensitivity competition of the bias and the nonlinear
interaction in spontaneous symmetry breaking}

The spontaneous symmetry breaking means that the symmetry is vulnerable to
the perturbation of the bias or the nonlinear interaction. It may be
worthwhile to compare the symmetry-breaking sensitivity to the bias and the
nonlinear interaction. As described in the paramagnetic-like and
antiferromagnetic-like symmetry patterns, let us remind that in the weak
linear coupling regime $g_{1}<g_{\mathrm{s}}$ the polarization $\langle
\sigma _{z}\rangle $ is more sensitive to the bias but responseless to the
nonlinear interaction within a threshold $g_{2c}$. We find this sensitivity
tendency is reversed in the strong linear coupling regime $g_{1}>g_{\mathrm{s%
}}$. It turns out that in this regime the symmetry breaking finds a higher
sensitivity to the nonlinear interaction than the bias. In Fig. \ref%
{Fig-sensativity}(a), we demonstrate that the symmetry breaking occurs
earlier in the nonlinear interaction (orange dashed line, $\epsilon =0$) in
the sense that the bias needs to have a relatively stronger strength (blue
dot-dashed line, $g_{2}=0$) to bring about the transition. Fig. \ref%
{Fig-sensativity}(b) shows the ratio of the critical-like strengths between
the bias and the nonlinear interaction. One sees the critical strength of
the bias is one or two orders larger than the nonlinear interaction.
Moreover, this ratio is growing with the linear coupling $g_{1}$.

One can see more clearly from the analytic boundary expressions %
\eqref{hzcII-0} and \eqref{g2cII-0}. We obtain the ratio between the
critical bias and nonlinear interaction
\begin{equation}
\frac{\left\vert \epsilon _{c}^{\mathrm{II}}\right\vert }{\left\vert
\widetilde{g}_{2c}^{\mathrm{II}}\right\vert }=\frac{\zeta ^{2}\overline{g}%
_{1}^{2}\Omega }{4g_{\mathrm{t}}}.
\end{equation}%
On the one hand, the low frequency contributes to the order difference as $%
g_{\mathrm{t}}=\omega /2.$ On the other hand, the ratio is proportional to $%
\overline{g}_{1}^{2}$ which grows parabolically with the strength of the
linear coupling. In addition, $\zeta $ starts for a small value at $g_{1}=g_{%
\mathrm{s}}$ and soon approaches to the value $1$ in the increase of $g_{1}$%
, which also contributes to the ratio growing at the beginning. Thus, unlike
in the regime below $g_{\mathrm{s}}$, the parity symmetry in the regime
beyond $g_{\mathrm{s}}$ is more sensitively broken by the perturbation of
nonlinear interaction than that of the bias, unless nearby $g_{\mathrm{s}}$.
This sensitivity priority of the nonlinear interaction comes from the
entanglement of the nonlinear interaction and the linear coupling, as will
be indicated by Eq.\eqref{bUpDown} in Section \ref{Sect-Mechanisms}.

\subsection{Tricriticality-(iv) induced by the interplay of the bias and the
nonlinear interaction at finite frequencies}

In Tricriticality-(iii) we have considered the bias and the nonlinear
interaction respectively. Now we should address how the tricritical point
and the fine structure of spontaneous symmetry breaking are affected by the
interplay of the bias and nonlinear interaction. In Fig. \ref%
{Fig-tricritical-2} we illustrate in panels (a-d) the phase diagrams by
variation of the bias in the presence of a fixed nonlinear interaction, and
in panels (e-h) the phase diagrams by variation of the nonlinear interaction
in the presence of a fixed bias. As one can see, besides the transition
boundaries I and II, two more boundaries appear as we mark by III and IV. As
expected, the onset of transition I can be seen by the start of increasing
in $\langle \sigma _{x}\rangle $,\ as shown in panels (a,e). Transitions II,
III and IV can be clearly observed in $\langle \sigma _{z}\rangle $ as
demonstrated in panels (b,f).

Although transition I is missed by $\langle \sigma _{z}\rangle $, all the
transitions I-IV leave some imprints in $\widetilde{x}_{\pm }$ as in panels
(c,d,g). The boundaries can also be all visualized by the peaks of the
susceptibility $d\widetilde{x}_{\pm }/dg_{1}$, as illustrated in panel (h).
For a fixed nonlinear interaction in panel (b-d), the boundary IV is tilted
upwards, with the critical bias increasing with the linear coupling. For a
fixed bias in panel (f-h), the boundary IV is tilted downwards, with the
critical nonlinear interaction\ decreasing with the linear coupling.

In Fig. \ref{Fig-tricritical-2} the crossing of the boundaries I and II
forms a first tricriticality around $g_{1}=1.2g_{\mathrm{s}},$\ which
actually is tricriticality-(iii) in the presence of only the bias or the
nonlinear interaction. Now in the presence of both the bias and the
nonlinear interaction, with the enhancement of the linear coupling the
boundaries II and III get closer to the tilted boundary IV and seem to form
a second tricritical-like point around $g_{1}=1.6g_{\mathrm{s}}$ which we
label by Tricriticality-(iv).

We extract in the leading order the analytic boundaries expressed by the
bias as a function of the linear coupling and the nonlinear interaction%
\begin{eqnarray}
\epsilon _{c}^{\mathrm{II}} &=&\frac{(1-t)\Omega }{4\delta _{c}\zeta }\exp [-%
\frac{\zeta ^{2}\overline{g}_{1}^{2}\Omega }{2\omega }]+\frac{1}{4}\zeta ^{2}%
\overline{g}_{1}^{2}\overline{g}_{2}\Omega ,  \label{hzC-II} \\
\epsilon _{c}^{\mathrm{III}} &=&-\frac{(1-t)\Omega }{4\delta _{c}\zeta }\exp
[-\frac{\zeta ^{2}\overline{g}_{1}^{2}\Omega }{2\omega }]+\frac{1}{4}\zeta
^{2}\overline{g}_{1}^{2}\overline{g}_{2}\Omega , \\
\epsilon _{c}^{\mathrm{IV}} &=&\frac{1}{4}\zeta ^{2}\overline{g}_{1}^{2}%
\overline{g}_{2}\Omega ,
\end{eqnarray}%
or tracked by the nonlinear interaction in variations of the linear coupling
and the bias
\begin{eqnarray}
\widetilde{g}_{2c}^{\mathrm{II}} &=&\frac{(1-t)g_{\mathrm{t}}}{\delta
_{c}\zeta ^{3}\overline{g}_{1}^{2}}\exp [-\frac{\zeta ^{2}\overline{g}%
_{1}^{2}\Omega }{2\omega }]+\frac{4\epsilon }{\zeta ^{2}\overline{g}%
_{1}^{2}\Omega }g_{\mathrm{t}}, \\
\widetilde{g}_{2c}^{\mathrm{III}} &=&-\frac{(1-t)g_{\mathrm{t}}}{\delta
_{c}\zeta ^{3}\overline{g}_{1}^{2}}\exp [-\frac{\zeta ^{2}\overline{g}%
_{1}^{2}\Omega }{2\omega }]+\frac{4\epsilon }{\zeta ^{2}\overline{g}%
_{1}^{2}\Omega }g_{\mathrm{t}}, \\
\widetilde{g}_{2c}^{\mathrm{IV}} &=&\frac{4\epsilon }{\zeta ^{2}\overline{g}%
_{1}^{2}\Omega }g_{\mathrm{t}},  \label{g2cIV}
\end{eqnarray}%
where $\overline{g}_{2}=\widetilde{g}_{2}/g_{\mathrm{t}}$. As shown in Fig. %
\ref{Fig-tricritical-2}(b,f) the analytic boundaries match the numerical
ones fairly well. \ We see that the interplay of the bias and the nonlinear
interaction contributes to the second term of the boundaries II and III, as
an additional term to \eqref{hzcII-0} and \eqref{g2cII-0}. Exactly speaking,
since the second term is equal to $\epsilon _{c}^{\mathrm{IV}}$ or $%
\widetilde{g}_{2c}^{\mathrm{IV}},$ the mathematical tricritical point (iv)
is at the infinity of linear coupling$.$ However in reality, although
boundary IV is actually comprised of boundaries II and III with $\widetilde{g%
}_{2c}^{\mathrm{IV}}$ or $\epsilon _{c}^{\mathrm{IV}}$\ as their center,
they are too close to be distinguished when the boundaries are tilted in the
regime of the strong linear coupling. Thus, effectively tricriticality-(iv)
appears at a finite value of the linear coupling.

\subsection{Tricriticality-(v) induced by frequency in the interplay of the
bias and the nonlinear interaction}
\label{Sect-TriCri-V}

%%%%%%%%%%%%%%%%%%%%%%%%%%%%%%%%%%%%%%%%%%%%%%%%%%%%%%%%%%%%%%%%%%%%%%%%%%%%%%%%%%%%%%%%%%%%%%%%%%
\begin{figure}[tbp]
\includegraphics[width=1.0\columnwidth]{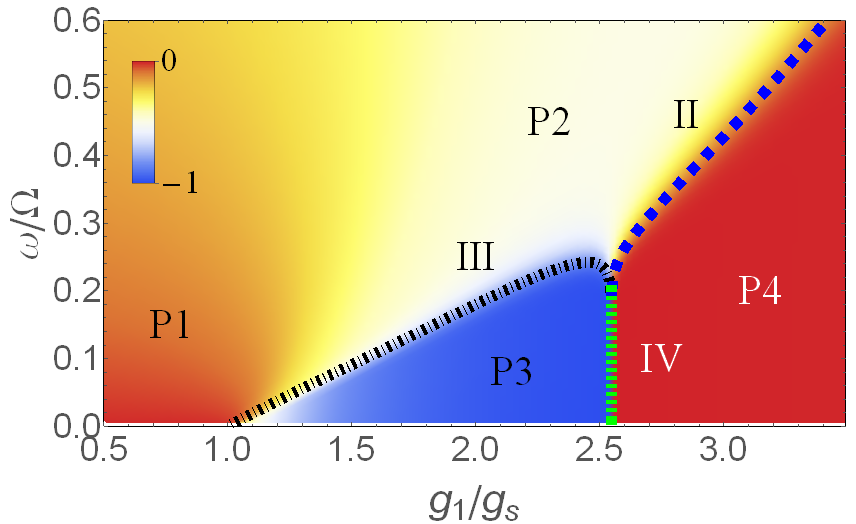}
\caption{(color online) \textit{Tricriticality-(v): the second
frequency-induced tricriticality.} Phase diagram of $\langle \hat{x}\rangle
_{+} /|x_{0,+}|$ in $g_{1}$-$\protect\omega $ plane at $\protect\epsilon %
=0.5\times 10^{-4}\Omega$ and $\log [g_{2}/g_{\mathrm{t}}]=-4.5$. P1, P2, P3
and P4 mark the different phases. The blue long-dashed, black dot-dashed and
green dashed lines are analytic boundaries II, III and IV. }
\label{Fig-Tricri-w-2}
\end{figure}
%%%%%%%%%%%%%%%%%%%%%%%%%%%%%%%%%%%%%%%%%%%%%%%%%%%%%%%%%%%%%%%%%%%%%%%%%%%%%%%%%%%%%%%%%%%%%%%%%%

Now let us come back to the frequency dimension. In Section \ref%
{Sect-finite-w}, we have seen that the frequency induces a tricritical point
in the respective presence of the bias or the nonlinear interaction. Now we
consider frequency effect in the presence of both the bias and the nonlinear
interaction. Imagine we are standing at the boundary IV in Fig. \ref%
{Fig-tricritical-2}, Eqs. \eqref{hzC-II}-\eqref{g2cIV} indicate that
increasing the frequency would open the gap between the boundary IV and the
boundaries II, III, thus inducing a tricritical-like behavior. We show such
a scenario by Fig. \ref{Fig-Tricri-w-2} in the $g_{1}$-$\omega $ plane. As
one can see, apart from the first frequency-induced tricritical point
(tricriticality-(ii) as afore-labeled) around $g_{1}=1.0g_{\mathrm{s}},$
another tricritical-like point appears around $g_{1}=2.5g_{\mathrm{s}}$
which is the location of $g_{1c}^{\mathrm{IV}}$ at a fixed bias $\epsilon
=0.0001\Omega $ and a nonlinear interaction $\log \left[ g_{2}/g_{\mathrm{t}}%
\right] =-4.5$. More generally, from Eq. \eqref{g2cIV} we extract the
location of the second frequency-induced tricriticality as
\begin{equation}
g_{1c}^{\mathrm{IV}}=g_{\mathrm{s}}\sqrt{\frac{2\epsilon +\sqrt{4\epsilon
^{2}+\overline{g}_{2}^{2}\Omega ^{2}}}{\Omega \overline{g}_{2}}}.
\label{g1cIV}
\end{equation}%
We label this tricriticality by (v). \ Exactly speaking, this tricritical
point is mathematically located at $\omega =0$, but effectively the
tricriticality seems to form at some finite frequency as boundaries II and
III are already too close to be distinguished at the finite frequency.

\subsection{Tendency for four successive transitions}
\label{Sect-4-transitions}

From the discussions in Section \ref{Sect-TriCri-low-w-limit}, we know that
in the low frequency limit there are at most two transitions in increasing $%
g_{1}$. The various situations for the occurrence of tricriticality
described above in Section \ref{SectionAll-Finte-w} demonstrate that finite
frequencies can lead to three transitions. Still, it might be possible to go
even further. A closer look at Fig. \ref{Fig-Tricri-w-2}, we can see the
boundaries II and III forms a dip shape around $g_{1}=2.5g_{\mathrm{s}}.$
The boundary III is actually a non-monotonic function of $g_{1}.$ Around $%
\omega =0.22\Omega ,$ in fact increasing $g_{1}$ goes across the boundary
III twice. \ Let us mark the different phases by P1, P2, P3, P4. In
increasing $g_{1}$ one starts with phase P1. After the first second-order
transition the system enters phase P2. Then the first time across boundary
III brings the system from phase P2 to phase P3. By the second time across
boundary III the system re-enters Phase P2. After the short re-entrance of
phase P2, the system transits to phase P4 through boundary II. Thus, in this
regime the system actually experiences four successive transitions, i.e.
transitions I, III, III and II, going through phases P1, P2, P3, P2, P4.
This tendency of the second additional transition indicates that finite
frequencies induces a subtle energy competition beyond the semiclassical
picture.

\section{Quadruple points and tetracriticality}\label{Sect-quadruple}

In last section we have seen that at finite frequencies the system can have
four phases P1, P2, P3 and P4 with three, even four transitions. We have
addressed a variety of situations in which tricriticality may occur. Since
we have four phases totally, one may wonder whether it is possible for all
the four phases to meet and form a quadruple point. We find this can happen
indeed. The possibility is indicated from the last transition point Eq. %
\eqref{hzcII-0} which, if the bias $\epsilon $ is being reduced, approaches
to the first transition $g_{1c}^{\mathrm{I}}=g_{\mathrm{s}}$\ in the low
frequency limit. This process of transition converging is shown in Fig. \ref%
{Fig-quadruple}, where we set $\epsilon =0.0005g_{\mathrm{t}}$ which is
proportional to the frequency. As one sees, all three transitions boundaries
finally collapses to one point at $g_{1}=g_{\mathrm{s}}$, thus forming a
quadruple point and a kind of tetracriticality. Again here, rather than four boundaries conventionally,
the tetracriticality here consists of three critical boundaries at finite frequencies and one critical point at zero frequency.
The critical point at zero frequency connects phases P1 and P4 directly.

\begin{figure}[tbp]
\includegraphics[width=1.0\columnwidth]{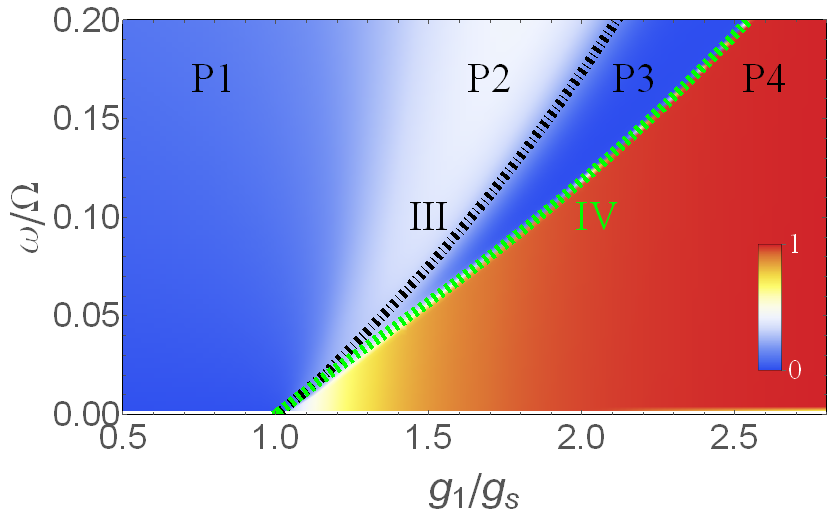}
\caption{(color online) \textit{Quadruple point and tetracriticality.} Phase diagram of $\langle
\hat{x}_{-}\rangle /|x_{0,-}|$ in $g_{1}$-$\protect\omega $ plane at $\log
[g_{2}/g_{\mathrm{t}}]=-4.5$ with $\protect\epsilon =0.0005g_{\mathrm{t}}$.
In low frequency limit the four phases P1, P2, P3 and P4 meet around $g_1=g_%
\mathrm{s}$, forming a quadruple point and a tetracriticality. }
\label{Fig-quadruple}
\end{figure}

The quadruple point is illustrated for a small value of $g_{2}$. One would
also get similar quadruple points in other values of $g_{2}$. The track by
varying $g_{2}$ continuously would yield a section of quadruple line along $%
\left\vert g_{1}\right\vert =g_{\mathrm{s}}\sqrt{1-\widetilde{g}_{2}^{2}/g_{%
\mathrm{t}}^{2}}$ which is actually the transition boundary %
\eqref{g1C-pure-g2} in the absence of the bias. Since the quadruple line is
parabolic in small values of $\widetilde{g}_{2}$, in weak nonlinear
interactions the quadruple points turn out to be around $\left\vert
g_{1}\right\vert =g_{\mathrm{s}}$ in the leading order, as we have seen in
the illustrated figure \ref{Fig-quadruple}.

%%%%%%%%%%%%%%%%%%%%%%%%%%%%%%%%%%%%%%%%%%%%%%%%%%%%%%%%%%%%%%%%%%%%%%%%%%%%%%%%%%%%%%%%%%%%%%%%%%
%%%%%%%%%%%%%%%%%%%%%%%%%%%%%%%%%%%%%%%%%%%%%%%%%%%%%%%%%%%%%%%%%%%%%%%%%%%%%%%%%%%%%%%%%%%%%%%%%%

\section{Changeovers of the wave function in the phase transitions}
\label{Sect-WaveFunction}

\begin{figure}[tbp]
\includegraphics[width=1.0\columnwidth]{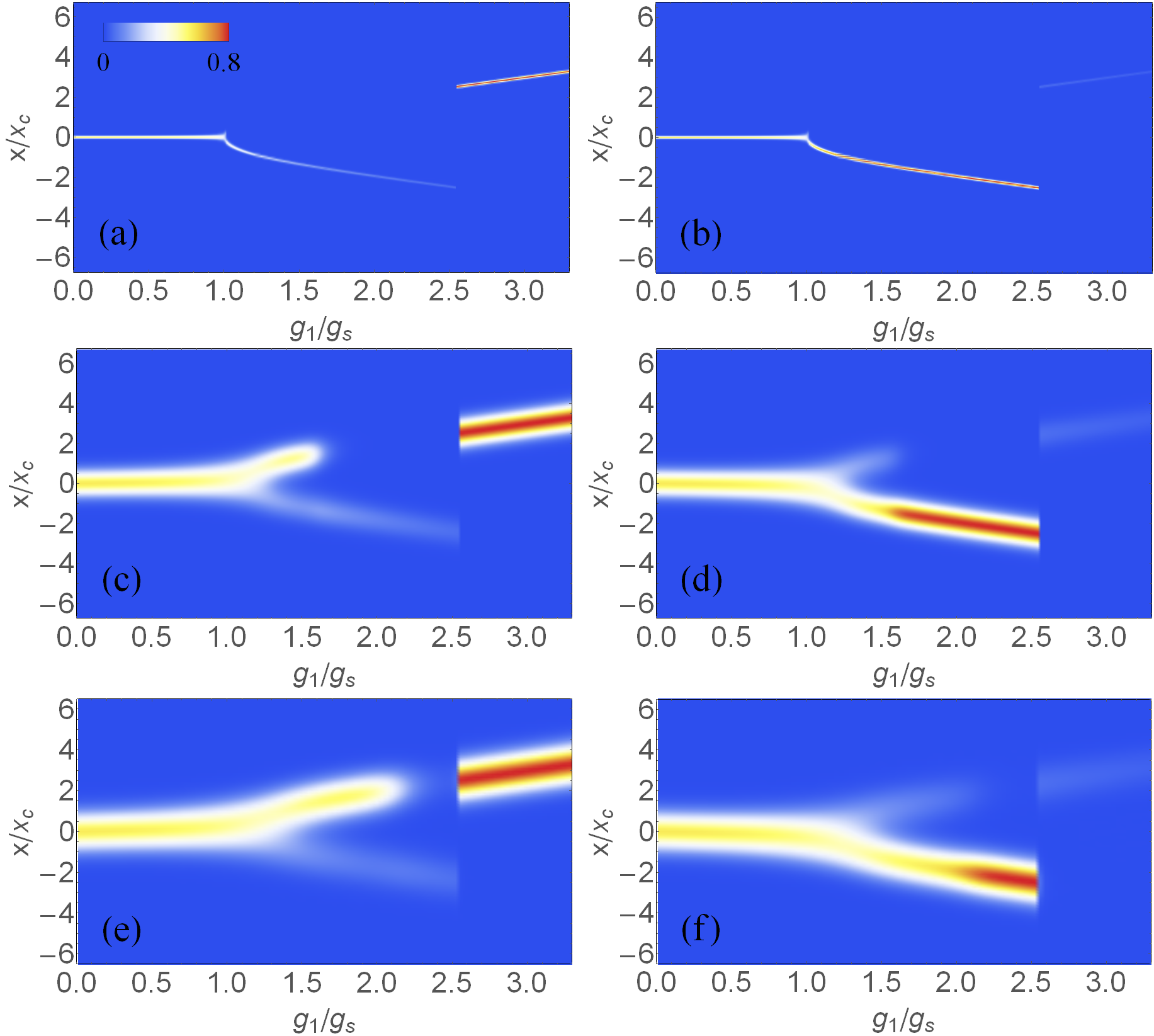}
\caption{(color online) Wave functions of spin-up (a,c,e) and spin-down
(b,d,f) components in phase transitions. (a,b) $\protect\omega=0.001\Omega$,
$\protect\epsilon=0.1g_\mathrm{t}$. (c,d) $\protect\omega=0.1\Omega$, $%
\protect\epsilon=0.001g_\mathrm{t}$. (e,f) $\protect\omega=0.2\Omega$, $%
\protect\epsilon=0.0005g_\mathrm{t}$. Here we fix $\log[g_2/g_\mathrm{t}]%
=-4.5$ and the effective spatial position is scaled by $x_c=\protect\sqrt{2}%
g_\mathrm{s}/\protect\omega.$ }
\label{Fig-wavefunction}
\end{figure}

To see the essential changes of quantum state in the transitions we shall
monitor the evolution of the wave function. In Fig.\ref{Fig-wavefunction} we
show the spin-up and spin-down components of\ the wave function that goes
through successive transitions in the variation of the linear coupling,
under fixed values of bias and nonlinear interaction. Panel (a,b) are in the
low frequency limit, while Panel (c)-(f) are finite-frequency\ cases. Note
different choices of frequency will change $g_{\mathrm{t}}$ which is taken
to be the strength reference of the nonlinear interaction as well as the
bias. Nevertheless by fixing two ratios $\widetilde{g}_{2}/g_{\mathrm{t}}$
and $\epsilon /\Omega $ we have the same transition point of the last
transition IV, around $g_{1}\sim 2.5g_{\mathrm{s}}$, which is the common one
in the different frequency illustrations, as indicated by Eq.\eqref{g1cIV}.

In the low frequency limit (illustrated by $\omega =0.001\Omega $) the wave
packet is very thin, just like a mass point of an effective particle, as one
sees from panels (a,b). Starting from $g_{1}=0$ till the first transition $%
g_{1}\sim 1.0g_{\mathrm{s}}$ the effective particle always stays at the
origin $x=0.$ Beyond the first transition it starts to go away from the
origin, and shifts to the other side at the next transition around $%
g_{1}\sim 2.5g_{\mathrm{s}}$.

At a finite frequency $\omega =0.1\Omega $ in panels (c,d) the wave packet
is obviously broadened, but still remaining in a single-branch structure and
staying around the origin before the first transition. After the first
transition the wave packet splits into two branches in both the spin
components, which is different from the low frequency limit. Strengthening
more the linear coupling $g_{1}$ triggers the second transition around $%
g_{1}\sim 1.6g_{\mathrm{s}}$ where one branch of the wave packet is broken.
In such a broken-branch state the wave packet on one side vanishes in both
spin components and all the weight goes to the branch on the other side.
Further increase of $g_{1}$ induces the third transition, around $g_{1}\sim
2.5g_{\mathrm{s}}$, which switches the broken-branch state from one side to
the other side. These three successive transitions correspond to the
boundaries I, III and IV in Fig. \ref{Fig-tricritical-2}(e,f) and Fig. \ref%
{Fig-Tricri-w-2}. Besides the different feature of the\ two-branch structure
after the first transition, the second transition is additional relative to
the\ low frequency limit. At a higher frequency $\omega =0.2\Omega $ in
panels (e,f)\ the second transition point moves to a stronger linear
coupling around $g_{1}\sim 2.2g_{\mathrm{s}}$. We also see that in the first
transition the splitting of the wave packet is continuous, which corresponds
to the second-order transition in $\langle \sigma _{x}\rangle $. The
changeover of the wave-function structure is discontinuous-like in the
second and third transitions, which matches the first-order-like transitions
in $\langle \sigma _{z}\rangle $.

The example is illustrated at small values of bias and nonlinear
interaction. It might be worth mentioning that at a fixed frequency a
stronger bias or nonlinear interaction can lead to a mixed quantum state,
i.e. one spin component in the two-branch state and the other spin component
in the broken-branch state. Further potential imbalance from the bias or
nonlinear interaction will finally drive both spin components into
broken-branch states.

%%%%%%%%%%%%%%%%%%%%%%%%%%%%%%%%%%%%%%%%%%%%%%%%%%%%%%%%%%%%%%%%%%%%%%%%%%%%%%%%%%%%%%%%%%%%%%%%%%
%%%%%%%%%%%%%%%%%%%%%%%%%%%%%%%%%%%%%%%%%%%%%%%%%%%%%%%%%%%%%%%%%%%%%%%%%%%%%%%%%%%%%%%%%%%%%%%%%%

\section{Mechanisms}
\label{Sect-Mechanisms}

In this section we should clarify the mechanisms underlying the various
patterns of symmetry breaking, the different orders of transitions and the
successive transitions in the tricritical picture. To facilitate the
understanding we rewrite bosonic mode in the model Hamiltonian in terms of
the quantum harmonic oscillator. By the transformation $a^{\dagger }=(\hat{x}%
-i\hat{p})/\sqrt{2}$, $a=(\hat{x}+i\hat{p})/\sqrt{2}$, we transfer to the
space of the effective position $\hat{x}$ and the momentum $\hat{p}$. Thus
the Hamiltonian takes the form
\begin{equation}
H=\sum_{\sigma _{z}=\pm }(h^{\sigma _{z}}|\sigma _{z}\rangle \langle \sigma
_{z}|+{\frac{\Omega }{2}}|\sigma _{z}\rangle \langle \overline{\sigma }_{z}|)
\end{equation}%
which is comprised of the effective free-particle part (the $h^{\sigma _{z}}$
term)\ and tunneling part (the $\Omega $ term).\ Here $\overline{\sigma }%
_{z}=-\sigma _{z}$ and $+$ ($-$) labels the up $\uparrow $ (down $\downarrow
$) spin. The effective free-particle Hamiltonian in the spin components can
be rearranged to be%
\begin{equation}
h^{\pm }=\omega \ (\frac{\hat{p}^{2}}{2m_{\pm }}+v_{\pm })+e_{0},\quad
v_{\pm }=v_{\pm }^{\mathrm{hp}}+b_{\pm }+b_{0}\mp \epsilon .
\label{h-v-UpDown}
\end{equation}%
where
\begin{eqnarray}
v_{\pm }^{\mathrm{hp}} &=&\frac{1}{2}m_{\pm }\varpi _{\pm }^{2}[x-x_{0,\pm
}]^{2}, \\
b_{\pm } &=&\pm \frac{\widetilde{g}_{2}^{\prime }g_{1}^{\prime 2}}{2(1-%
\widetilde{g}_{2}^{\prime 2})},  \label{bUpDown} \\
b_{0} &=&-g_{1}^{\prime 2}/[2(1-\widetilde{g}_{2}^{\prime 2})].
\end{eqnarray}%
We have defined $g_{1}^{\prime }=\sqrt{2}g_{1}/\omega $, $g_{2}^{\prime
}=2g_{2}/\omega $ and \ $e_{0}=-\omega /2$. Here $m_{\pm }=\left( 1\mp
g_{2}^{\prime }\pm \chi g_{2}^{\prime }\right) ^{-1}$ is the effective mass,
$\varpi _{\pm }=[\left( 1\pm \chi g_{2}^{\prime }\right) ^{2}-g_{2}^{\prime
2}]^{1/2}$ is frequency renormalization. The $x_{0,\pm }=\mp g_{1}^{\prime
}/\left( 1\mp \widetilde{g}_{2}^{\prime }\right) $ is the potential
displacement for the potential minimum shifting horizontally from the
origin, while $b_{0}$ is the vertical shift which is both downward for the
two spin components. In this picture we see the different roles played by
the physical parameters of the model: the linear coupling $g_{1}$ separates
the potentials of the two spin components, the bias $\epsilon $\ shifts the
potentials downwards or upwards oppositely for the two spin components,
while the nonlinear interaction $g_{2}$ not only leads to asymmetry in
frequency $\varpi _{\pm }$ and potential displacement $x_{0,\pm }$ but also
results in vertical potential difference $b_{\pm }.$

\subsection{Semiclassical picture for the various patterns of symmetry
breaking}
\label{Sect-Semiclassical}

%%%%%%%%%%%%%%%%%%%%%%%%%%%%%%%%%%%%%%%%%%%%%%%%%%%%%%%%%%%%%%%%%%%%%%%%%%%%%%%%%%%%%%%%%%%%%%%%%%
\begin{figure}[tbp]
\includegraphics[width=1.0\columnwidth]{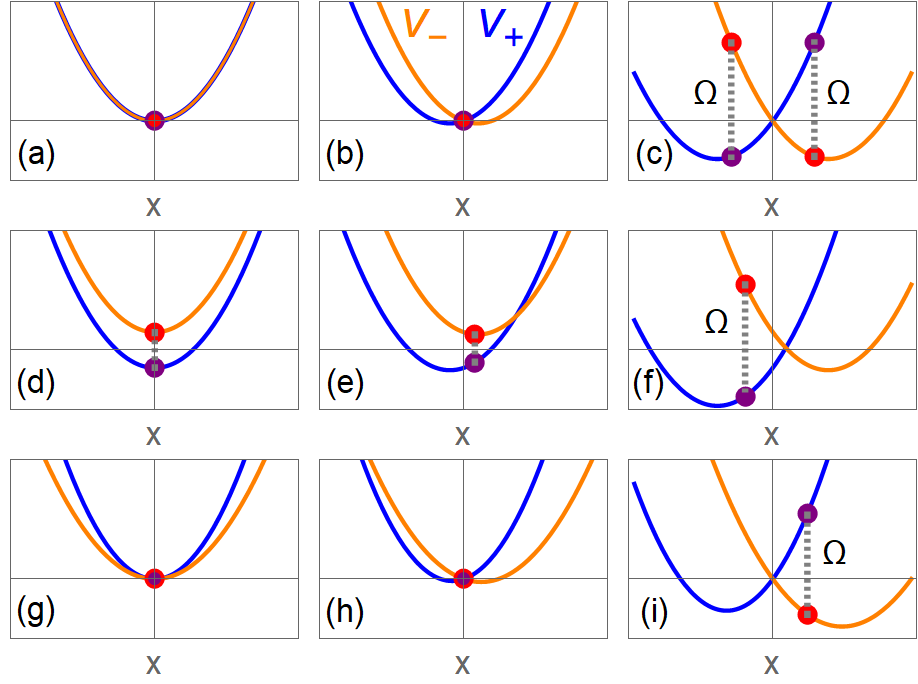}
\caption{(color online) \textit{Semiclassical mechanisms for the different
patterns of symmetry breaking.} The potentials for spin-up (blue) and
spin-down (orange) components for (a-c) $\protect\epsilon =0,g_{2}=0$, (d-f)
$\protect\epsilon \neq 0,g_{2}=0$ and (g-i) $g_{2}\neq 0, \protect\epsilon %
=0 $. The linear coupling regimes are (a,d,g) $g_{1}=0$, (b,e,h) $%
g_{1}<g_{s} $ and (c,f,i) $g_{1}>g_{s}$. The dots mark the effective
semi-classical particle positions in the spin-up (purple) and spin-down
(Red) potentials. }
\label{Fig-potentials}
\end{figure}
%%%%%%%%%%%%%%%%%%%%%%%%%%%%%%%%%%%%%%%%%%%%%%%%%%%%%%%%%%%%%%%%%%%%%%%%%%%%%%%%%%%%%%%%%%%%%%%%%%

The phase transitions of the quantum Rabi model occurs at low frequencies.
The ground-state wave function basically can be decomposed into ground
states of quantum harmonic oscillators with displacement and frequency
renormalizations \cite{Ying2015}. The wave-packet size is of order $1$ in
the afore-presented dimensionless formalism. The potential size at phase
transitions can be estimated by $x_{0,\pm },$\ being of order $g_{s}^{\prime
}=\sqrt{2}g_{s}/\omega \propto \sqrt{\Omega /\omega }$. Thus the ratio
between the wave-packet size and the potential size is of order $\sqrt{%
\omega /\Omega }$ which becomes smaller at a lower frequency. In the low
frequency limit, $\omega /\Omega \rightarrow 0$, with the wave-packet size
relatively negligible, one can regard the effective particle as a classical
mass point, as we have seen in Fig.\ref{Fig-wavefunction}(a,b). On the other
hand we keep the leading tunneling effect in the spin\ space. In such a
semiclassical consideration, the ground state is motionless with $%
p\rightarrow 0$, thus the phase transitions and the system properties are
decided by the competition of the potential $v_{\pm }$ and the tunneling $%
\Omega $.

The various patterns of symmetry breaking in the low frequency limit can be
readily explained in such semiclassical picture. In Fig. \ref{Fig-potentials}%
, according to different patterns of symmetry breaking we plot the
potentials $v_{+}$(blue) for up spin and $v_{-}$(orange) for down spin. The
purple (spin up) \ and red (spin down) dots mark the positions of the
effective mass point and the spin tunneling is indicated by he gray dashed
lines.

Fig.\ref{Fig-potentials}(a-c) present the situation of the conventional
quantum Rabi model, in the absence of the bias and the nonlinear
interaction. Starting from the zero linear coupling $g_{1}=0$ in panel (a),
the spin potentials are identical, with the effective particle staying at
the origin where the potential minima are located. The increase of $g_{1}$
separates the potentials horizontally by $x_{0,\pm }$, as indicated in panel
(b). However, with a linear coupling below $g_{s}$, the effective particle
in the two spin components does not follow the potential separation but
remains at the origin instead. This is because moving away from the origin
would lose the negative tunneling energy due to the unequal spin weights in
the potential difference, while staying at the origin keeps the maximum
tunneling energy due to equal spin weights in the degenerate potentials.
Increasing $g_{1}$ beyond the critical point,\ the downward potential shift
by $b_{0}$ enlarges the potential difference between the bottom and the
origin as in panel (c), so that moving toward the potential bottom will gain
more potential energy than the tunneling energy. Therefore the transition
occurs and the particle leaves the origin. Note that either before or after
the transition the spin distributions are spatially symmetric around the
origin and the weights remain equal under spin exchange, thus the parity
symmetry is preserved throughout.

Fig.\ref{Fig-potentials}(d-f) denote the situation of adding a bias to the
linear coupling. A bias separates vertically the potentials of the up and
down spins at $g_{1}=0$, as in panel (d), which breaks the spin balance and
the parity symmetry from the beginning, thus being paramagnetic-like in
polarization. In weak linear coupling regime, the bias moves the potential
crossing point away from the origin which breaks the space inversion
symmetry of the potential. On the other hand, the crossing point is moving
to a higher potential which is not energetically favorable. So the parity
symmetry is broken in both the spatial and spin parts. In a strong linear
coupling beyond the critical point, as in panel (f), any strength of the
bias will break the two-side balance maintained by the linear coupling in
panel (c), thus a spontaneous symmetry breaking occurs. Note the state on
each side is polarized due to the finite difference in spin-up and spin-down
energy. Before the spontaneous symmetry breaking, the polarization or spin
expectation $\langle \sigma _{z}\rangle $ cancels between the two sides.
After the\ spontaneous symmetry breaking, without the two side cancellation,
the polarization jumps to a finite value.

Fig.\ref{Fig-potentials}(g-i) show the situation of adding a nonlinear
interaction to the linear coupling. The nonlinear interaction makes the
frequency asymmetric between the up and down spins as in panel (g) and also
shifts the spins in vertically opposite directions as in panel (h). However
the potential crossing always keeps invariant at the origin. Thus the parity
is well preserved even in the presence of a finite nonlinear interaction.
Note that the vertical spin-dependent shift $b_{\pm }$ in Eq. (\ref{bUpDown}%
) has an entangled form of the linear coupling $g_{1}$ and the nonlinear
interaction $g_{2}$, increasing the nonlinear interaction at a fixed linear
coupling will enlarge the vertical potential difference between the two spin
directions. This vertical potential difference will finally surpass the
tunneling energy at the origin and lead to symmetry breaking with a first
order transition. So the polarization behavior is ferromagnetic-like. In a
strong linear coupling beyond the $g_{s}$, also a tiny strength of nonlinear
interaction will break the balance on the two sides in panel (c), leading to
a spontaneous symmetry breaking from panel (c) to panel (i).

From the basic competitions discussed in the above one can also understand
similarly the other mixed patterns of symmetry breaking. For the transition
orders we will present some explanations from the view of the variational
energy later on in Section \ref{Sect-E-SemiClasiscal}.

\subsection{Scaling of the Stark term}
\label{Sect-Scaling-Stark}

%%%%%%%%%%%%%%%%%%%%%%%%%%%%%%%%%%%%%%%%%%%%%%%%%%%%%%%%%%%%%%%%%%%%%%%%%%%%%%%%%%%%%%%%%%%%%%%%%%
\begin{figure}[tbp]
\includegraphics[width=0.85\columnwidth]{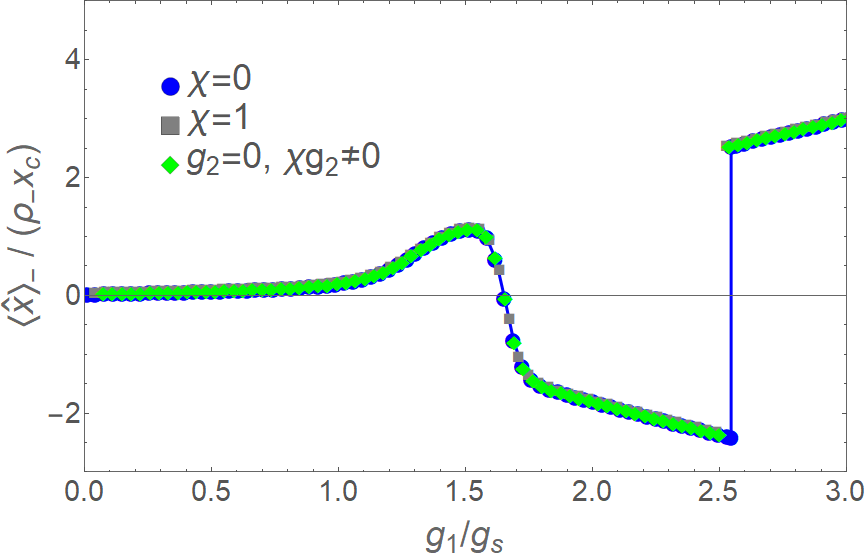}
\caption{(color online) \textit{Scaling of the Stark term.} $\langle\hat{x}%
_{-}\rangle /(\protect\rho_{-} x_c)$ versus $g_1$ with different Stark
couplings $\protect\chi =0$ (blue dots), $\protect\chi =1$ (gray squares)
and $g_2=0,\protect\chi g_2 \neq 0$ (green diamonds) at a same value of $%
\widetilde{g}_2=(1+\protect\chi)g_2$ . Here $\protect\omega=0.1\Omega$, $%
\protect\epsilon=0.001g_\mathrm{t}$ and $\log[\widetilde{g}_2/g_\mathrm{t}]
=-4.5$. }
\label{Fig-Stark-Scaling}
\end{figure}
%%%%%%%%%%%%%%%%%%%%%%%%%%%%%%%%%%%%%%%%%%%%%%%%%%%%%%%%%%%%%%%%%%%%%%%%%%%%%%%%%%%%%%%%%%%%%%%%%%

As mentioned around Eq.\eqref{scaled-g2}, the properties with the Stark term
are similar by included the scaling factor, unless the frequency is high. We
illustrate the scaling in Fig.\ref{Fig-Stark-Scaling} where it is shown that
different Stark couplings under a fixed value of $\widetilde{g}%
_2=(1+\lambda)g_2$ have the same spin expectation and the same successive
transition points (around $g_1/g_\mathrm{s}\sim 1.0,\ 1.6,\ 2.6$). This
scaling can be simply understood from the semiclassical picture
afore-formulated. In fact, from Eq.\eqref{h-v-UpDown} we have seen that the
potential displacement $x_{0,\pm },$ the effective bias $b_{\pm }$ and and
the uniform shift $b_{0}$ are all functions of $\widetilde{g}_{2}^{\prime
}=\left( 1+\chi \right) g_{2}^{\prime }.$ It should be noted that, although
the effective mass $m_{\pm }$ and $\varpi _{\pm }$ respectively are not
functions of $\widetilde{g}_{2}^{\prime },$ their joint contribution in $%
v_{\pm }^{\mathrm{hp}}$ is still a function of $\widetilde{g}_{2}^{\prime }$
as
\begin{equation}
m_{\pm }\varpi _{\pm }^{2}=\frac{[\left( 1\pm \chi g_{2}^{\prime }\right)
^{2}-g_{2}^{\prime 2}]}{\left( 1\mp g_{2}^{\prime }\pm \chi g_{2}^{\prime
}\right) }=\left( 1\pm \widetilde{g}_{2}^{\prime }\right) .
\end{equation}%
Namely, except for the kinetic term neglected in the semiclassical picture
in the low frequency limit, all contributions of the Stark-like term to $%
v_{\pm }$ can be scaled into a function of $\widetilde{g}_{2}^{\prime }.$
Thus, one will get the same phase diagrams for the presence of the
Stark-like term by the scaling factor $\left( 1+\chi \right) $.

\subsection{Semiclassical energy competition for the different orders of
phase transitions}
\label{Sect-E-SemiClasiscal}

%%%%%%%%%%%%%%%%%%%%%%%%%%%%%%%%%%%%%%%%%%%%%%%%%%%%%%%%%%%%%%%%%%%%%%%%%%%%%%%%%%%%%%%%%%%%%%%%%%
\begin{figure}[tbp]
\includegraphics[width=1.0%
\columnwidth]{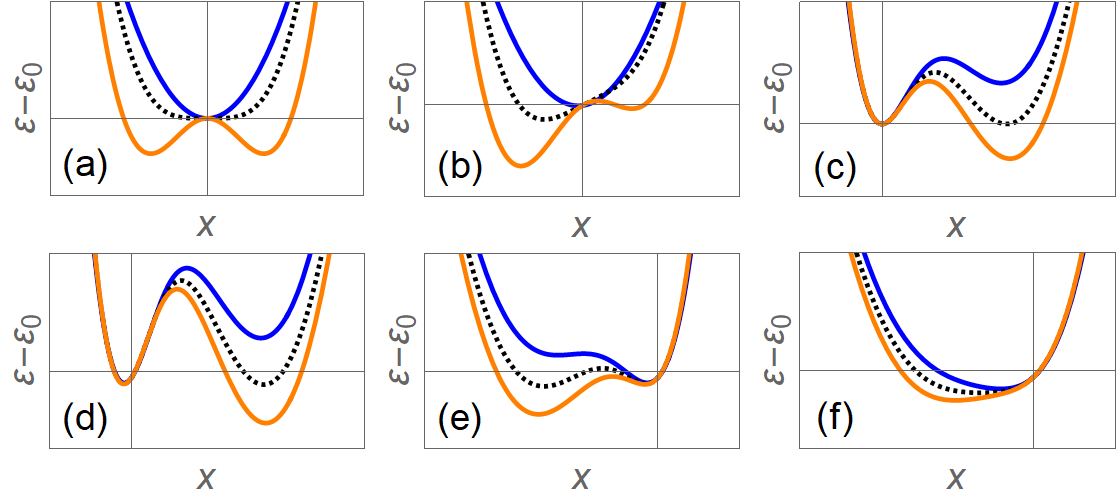}
\caption{(color online) \textit{Energy competitions and transition orders.}
Semiclassical variational energy $\protect\varepsilon $ before (blue solid
lines), at (black dotted lines) and after (orange solid lines) transitions,
with respect to the effective particle position $x$ for (a) $\protect%
\epsilon =0$ and $g_{2}=0$, (b) $\protect\epsilon \neq 0$ and $g_{2}=0$, (c)
$\protect\epsilon =0$ and $g_{2}\neq 0$, (d) $\protect\epsilon \neq 0$ and $%
g_{2}>0$, (e) $\protect\epsilon \neq 0$ and $g_{2}<0$ nearby $g_{2}=g_{t}$,
(f) $\protect\epsilon \neq 0$ and $g_{2}<0$ nearby $g_{2}=0$. Here $\protect%
\varepsilon _{0}=-( \protect\omega +\Omega )/2$.}
\label{Fig-variational-E}
\end{figure}
%%%%%%%%%%%%%%%%%%%%%%%%%%%%%%%%%%%%%%%%%%%%%%%%%%%%%%%%%%%%%%%%%%%%%%%%%%%%%%%%%%%%%%%%%%%%%%%%%%

We can gain more insights from the total energy competition. The variational
energy in semiclassical picture can be formulated in the following
eigenequation of matrix form
\begin{equation}
\left(
\begin{array}{cc}
\varepsilon _{+} & -\frac{\Omega }{2} \\
-\frac{\Omega }{2} & \varepsilon _{-}%
\end{array}%
\right) \left(
\begin{array}{c}
\beta ^{+} \\
\alpha ^{-}%
\end{array}%
\right) =\varepsilon \left(
\begin{array}{c}
\beta ^{+} \\
\alpha ^{-}%
\end{array}%
\right)  \label{H-matrix-semiclassical}
\end{equation}%
where $\varepsilon _{\pm }=\omega \ v_{\pm }+\varepsilon _{0}.$ The
eigenenergy for the ground state is determined by
\begin{equation}
\varepsilon =\frac{1}{2}\left[ \left( \varepsilon _{+}+\varepsilon
_{\_}\right) -\sqrt{\left( \varepsilon _{+}-\varepsilon _{\_}\right)
^{2}+\Omega ^{2}}\right]
\end{equation}%
which should be minimized with respect to $x$ as $v_{\pm }$ is position
dependent.

In Fig.\ref{Fig-variational-E} we illustrate the variational energy as a
function of $x$ before the transition (blue solid lines), at the transition
(black dashed lines) and after the transition (orange solid line) in
different situations. Panel (a) presents the case of the conventional
quantum Rabi model without the bias and the nonlinear interaction. Before
the transition the energy minimum is located at the origin, after the
transition the origin becomes an unstable saddle point while the ground
state lies in the formed two symmetric minima which are moving away from the
origin. At the transition point $g_{1}=g_{\mathrm{s}}$ the minimum bottom
becomes flat with a vanishing second derivation $\partial ^{2}\varepsilon
/\partial x^{2}=0$. Although the transition turns the minimum number from
one to two, this transition is continuous as two minimum positions separate
continuously from the origin.

The presence of the bias breaks the\ symmetry in the energy profile in any
regime of the linear coupling, as illustrated in Fig.\ref{Fig-variational-E}%
(b). The profile difference of single minimum and double minimum in energy
leads to different response to the bias before and after the transition.
Before the transition point $g_{\mathrm{s}}$ the energy has no competition
as the single minimum is the only choice. With the bias this single minimum
moves gradually away from the origin. After the transition, there are two
minima which are degenerate in the absence of the bias. Any tiny strength of
bias will immediately break the symmetry and raise the degeneracy. Changing
the sign of the bias the ground state will shift from one side of the
minimum to the other side. Either the bias opening or sign change will lead
to an abrupt jump in polarization, leading to a discontinuous first-order
transition.

The scenario of energy competition is different in the presence of nonlinear
interaction, as demonstrated in Fig.\ref{Fig-variational-E}(c). There are
two local energy minima both before and after the transition\ (here the
transition moves from $g_{\mathrm{s}}$ to $g_{1c}$ in Eq.\eqref{g1C-pure-g2}%
), one at the origin, the other away from the origin. Before the transition,
the ground state lies in the minimum at the origin while the other local
minimum is higher in energy. At the transition the higher minimum is lowered
to get degenerate with the one at the origin. After the transition, the
energy preference gets reversed and the ground state turns to the lower
minimum away from the origin. Note that, in a sharp contrast to the
continuous variation of the minimum position in panel (a), the transition
here in panel (c) is companied with a sudden shift of minimum position. This
discontinuous shift of minimum position results in the first-order
transition. Conventionally continuous/discontinuous transitions refer to
continuity of different-order energy derivatives with the respect to the
system parameter, here the\ continuous/discontinuous variation of minimum
position provides another angle of view from the aspect of the
variational-energy structure.

In the presence of both the bias and the nonlinear interaction, there are
three situations which should be distinguish. Fig.\ref{Fig-variational-E}(d)
shows the first case in which the bias $\epsilon $ and the nonlinear
interaction $g_{2}$ have the the same sign. In his case the bias pushes the
minimum at the origin away to the opposite side of the higher minimum. In
this case the transition also is discontinuous, similar to panel (c), which
accounts for the first-order boundary in the positive-$g_{2}$ regime of Fig. %
\ref{Fig-SemiClassicDiagrams}(d). This first order transition boundary
covers all range of the linear coupling $g_{1}$. Fig.\ref{Fig-variational-E}%
(e) shows the second case with the sign of $g_{2}$ opposite to $\epsilon $
and the amplitude of $g_{2}$ closer to $g_{\mathrm{t}}$. In this case the
two energy minima are located on the same side but still far away enough to
have a barrier between them. Thus the transition also has a discontinuous
shift of the minimum position, which corresponds to the first-order boundary
arc in the negative-$g_{2}$ regime of Fig. \ref{Fig-SemiClassicDiagrams}(d).
The third situation shown in Fig.\ref{Fig-variational-E}(f) still has
opposite signs of $g_{2}$ and $\epsilon $ but with a small amplitude of $%
g_{2}$. In this case the two local energy minima are too close to have a
barrier to separate them explicitly. Although the minimum position may have
a quick shift but the variation is continuous. Thus the first order
transition is softened, being second-order like or even fading away. This
corresponds to the regime above the arc boundary in Fig. \ref%
{Fig-SemiClassicDiagrams}(d) where the first-order boundary disappears.

\subsection{Full-quantum-mechanical effect for the novel successive
transitions and tricriticalities}
\label{Sect-tricritical-mechanism}

%%%%%%%%%%%%%%%%%%%%%%%%%%%%%%%%%%%%%%%%%%%%%%%%%%%%%%%%%%%%%%%%%%%%%%%%%%%%%%%%%%%%%%%%%%%%%%%%%%
\begin{figure}[tbp]
\includegraphics[width=1.0\columnwidth]{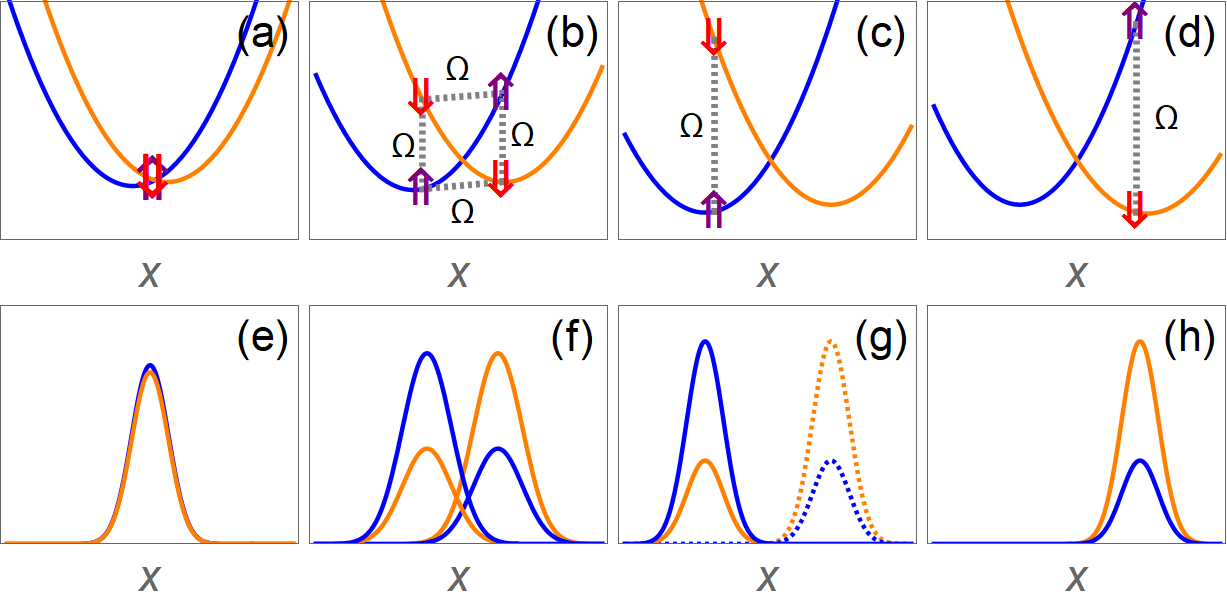}
\caption{(color online) \textit{Full-quantum-mechanical mechanisms for
additional transition in successive transitions.} (a-d) Effective potentials
for the spin-up (blue) and spin-down (orange) components, the arrows
represents the spins and gray dashed lines denotes the tunneling channels.
(e-h) Schematic decomposed wave functions in the spin-up (blue) and
spin-down (orange) components. The dashed lines in (g) show the vanishing
left-right wave-packet overlap and indicate the disappearing wave packets. }
\label{Fig-tricritical-mechanism}
\end{figure}
%%%%%%%%%%%%%%%%%%%%%%%%%%%%%%%%%%%%%%%%%%%%%%%%%%%%%%%%%%%%%%%%%%%%%%%%%%%%%%%%%%%%%%%%%%%%%%%%%%

In the afore-discussed semiclassical picture there is no spatial structure
of wave function or probability distribution over the effective spatial
space. This simplification will miss some physics that becomes important at
finite frequencies. Indeed, as described in Section \ref{SectionAll-Finte-w}%
, the novel successive transitions and tricriticalities emerge at a finite
frequencies, which cannot be captured by the semiclassical picture. To
understand these novel phenomena we shall fall back on a
full-quantum-mechanical picture. To include all the quantum states in one
example we follow the wave function evolution in Fig.\ref{Fig-wavefunction}%
(c,d) where there are four quantum states. Accordingly, in Fig. \ref%
{Fig-tricritical-mechanism}, we sketch the spin potentials (upper panels)
and the wave-function profiles (lower panels). The wave function is
decomposed into left and right wave packets, as analyzed in a
polaron-antipolaron picture \cite{Ying2015}, due to the barrier indicated in
Fig.\ref{Fig-variational-E}. Each wave packet is represented by a displaced
ground state of quantum harmonic oscillator \cite{Ying2015,Bera2014Polaron}
and the heights indicate the weights.

There are three transitions in the illustrated case, going through
transitions I, III and IV in Fig.\ref{Fig-tricritical-2}(e,f). Before the
first transition, the tunneling energy is dominating. As in Fig.\ref%
{Fig-tricritical-mechanism}(a,e), the single wave packets in both spins
reside around the origin where the potential crossing point is located. The
degeneracy at the crossing point yield equal weights of the two spin
components. Both the single-wave-packet profiles and equal spin weights help
to gain a maximum tunneling energy. The equal spin-component weight and the
full overlapping yield a vanishing spin expectation $\langle \sigma
_{z}\rangle $ and a saturation of $\langle \sigma _{x}\rangle .$

Transition-I: \ Increasing the linear coupling separates the potentials more
and lower the potential bottoms, so that the potential energy comes to
compete with the tunneling energy. After the first transition, as in Fig. %
\ref{Fig-tricritical-mechanism}(b,f), the wave function splits into four
wave packets. Differently from the semiclassical picture there are now four
channels of tunneling.\ The left-right tunneling arises due to the
left-right overlaps of the wave packets, while the semiclassical particle
has no such left-right overlap in any case. These left-right tunneling
channels come to play an important role to balance potential asymmetry
caused by the bias or the nonlinear interaction. Note in such a four-channel
state, the polarizations of the two sides are canceling each other so that $%
\langle \sigma _{z}\rangle $ still remains almost vanishing at the presence
of a weak bias and nonlinear interaction. Therefore, $\langle \sigma
_{z}\rangle $ does not have an obvious change across the first transition
thus the first transition leaves little imprint in $\langle \sigma
_{z}\rangle $. On the other hand, the separated wave packets are moving away
from the origin, the potential difference leads to unequal weights of the
two spin components on each side. This weight difference lead to the
reduction of spin flipping amplitude thus $\langle \sigma _{x}\rangle $ is
decreasing in strength. As a result, $\langle \sigma _{x}\rangle $ is
sensitive to the first transition and exhibits a critical behavior of
second-order transition.

Transition-III: Further increase of $g_{1}$ will separate the wave packets
more so that the left-right overlap becomes vanishing, as indicated in Fig. %
\ref{Fig-tricritical-mechanism}(g). Thus the left-right channels of
tunneling in a vanishing strength cannot balance the potential asymmetry any
more. As a result, the wave packets on the higher-potential side disappear
and the second transition occurs. Note that at a higher frequency would have
wider wave packets, thus the left-right overlap survive till larger $g_{1}$
and the transition occurs later. After this transition, the left-right
cancellation does not exist in the one-side state so that $\langle \sigma
_{z}\rangle $ jumps from a vanishing value to a finite value. Consequently
this transition can find a clear signal in $\langle \sigma _{z}\rangle $. \
On the other hand, the state on the two sides have a similar amplitude of
difference in the weights for the spin components. Note the strength of $%
\langle \sigma _{x}\rangle $ is decided by the weight difference of the two
spins no matter which spin component has more weight. Thus $\langle \sigma
_{x}\rangle $ does not respond to this transition unless the potential
asymmetry is large in the presence of strong bias and nonlinear interaction.
Transition II has the same nature as transition III, although not present in
the example of Fig.\ref{Fig-tricritical-mechanism}.

Transition-IV: An even larger $g_{1}$ will enhance much the entangled
effective bias $b_{\pm }$ which is proportional to $g_{1}^{2}$. This
enhanced bias will surpass the system bias $\epsilon $ which is originally
stronger in small-$g_{1}$ regime. This strength reversion of the two
competing biases triggers transition IV. In principle, at the reversion the
system should return to the four-wavepacket state. However the left-right
overlap is too small to maintain four-wavepacket state long enough to open a
phase, unless the frequency is higher to get more-broadened wave packets.
Hence, transition IV simply appears as one sharp transition. At a higher
frequency the wave packets could be more more-broadened so that some
left-right overlap could still remain, in such a situation Transition-IV
could be bifurcated into two close transitions as mentioned for the tendency
of four successive transitions in Section \ref{Sect-4-transitions}. Since
the state shifts from one side to the other, the sign of $\langle \sigma
_{z}\rangle $ get reversed so that $\langle \sigma _{z}\rangle $ exhibits a
first-order change at this transition. In the same reason as in
Transition-III $\langle \sigma _{x}\rangle $ still shows no sign at
Transition-IV.

From the above understanding we see that $\langle \sigma _{x}\rangle $ can
be sensitive to measure the first transition and $\langle \sigma _{z}\rangle
$ is useful to track all the other transitions. The spin-filtered quantity $%
\langle {a^{\dag }+a\rangle }_{\pm }/\rho _{\pm }$ is the spin displacement
in our picture, i.e, the effective wave-packet position in each spin
component. So it is naturally sensitive to the side shifting in transitions
II,III,IV. Moreover, in the four-wavepacket state after transition I, each
spin component has imbalanced weights of the left-side wave packet and the
right-side wave packet, due to the potential difference within $v_{+}$ or $%
v_{-}$ as shown in Fig.\ref{Fig-tricritical-mechanism}(b). Indeed $\langle {%
a^{\dag }+a\rangle }_{\pm }/\rho _{\pm }$ reflects the effective mass center
of each spin component, which is moving away from the origin after
transition I. In consequence, $\langle {a^{\dag }+a\rangle }_{\pm }/\rho
_{\pm }$ is also responding to transition I, thus useful to detect all
transitions simultaneously.

\section{Finding analytic boundaries}
\label{Sect-Analytic}

In previous sections \ref{Sect-PhaseDiagram-low-w}-\ref{Sect-quadruple} we
have given the analytic phase boundaries in the description of the phase
diagrams. In Sections \ref{Sect-WaveFunction} and \ref{Sect-Mechanisms} we
have got a basic understanding of the different transitions from the wave
function changeovers and energy competitions. This would facilitate the
finding of analytic boundaries. Now we try to provide some brief derivations
for the analytic phase boundaries.

\subsection{Analytic boundaries in low frequency limit}
\label{Sect-Analytic-low-w}

%%%%%%%%%%%%%%%%%%%%%%%%%%%%%%%%%%%%%%%%%%%%%%%%%%%%%%%%%%%%%%%%%%%%%%%%%%%%%%%%%%%%%%%%%%%%%%%%%%
\begin{figure}[tbp]
\includegraphics[width=0.8\columnwidth]{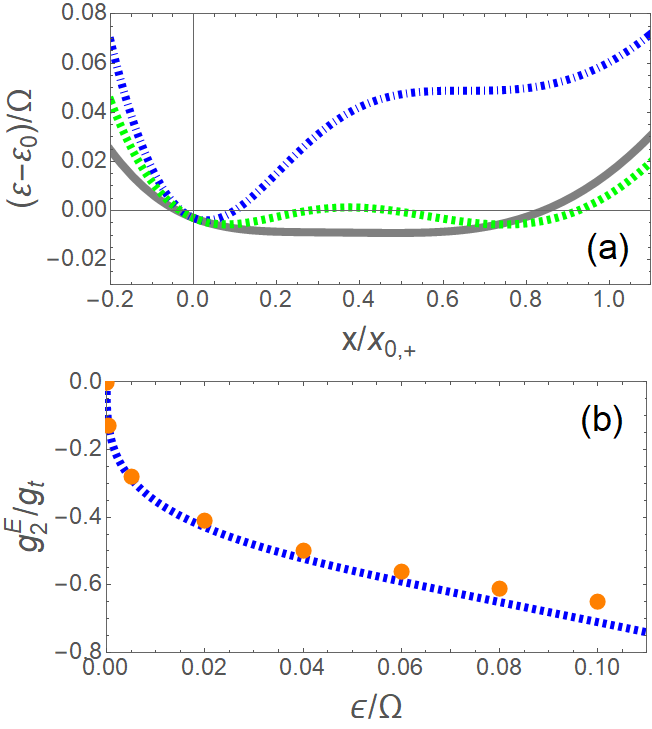}
\caption{(color online) \textit{Saddle point flattening and ending points of
the arc boundary.} (a) Semiclassical variational energy at the end of the
arc boundary $g_1=0.763g_\mathrm{s}$, $g_2=-0.554g_\mathrm{t}$(gray solid),
in the arc $g_1=0.7g_\mathrm{s}$, $g_2=-0.665g_\mathrm{t}$ (green dashed)
and at an infection point $g_1=0.5g_\mathrm{s}$, $g_2=-0.81g_\mathrm{t}$
(blue dot-dashed). Here $\protect\epsilon=10g_\mathrm{t}$ and $\protect\omega%
=0.001\Omega$. (b) Critical value $g_2^\mathrm{E}$ at the end of the arc
boundary versus the bias $\protect\epsilon$ from numerics (dots) and the
analytic result (blue dashed). }
\label{Fig-Saddle-flat}
\end{figure}
%%%%%%%%%%%%%%%%%%%%%%%%%%%%%%%%%%%%%%%%%%%%%%%%%%%%%%%%%%%%%%%%%%%%%%%%%%%%%%%%%%%%%%%%%%%%%%%%%%

With the clarifications of the mechanisms for all the transitions, we can
extract the analytic phase boundaries. In the low frequency limit, the
boundary can be obtained from the semiclassical picture. The energy minima
can be available by minimization of the variational energy $\varepsilon $
with respect to the position,
\begin{equation}
\frac{\partial }{\partial x}\varepsilon \left( x\right) =0
\end{equation}%
which gives three roots $x_{\mathrm{R}}$, $x_{\mathrm{S}}$, $x_{\mathrm{L}}$%
. The root $x_{\mathrm{S}}$ between the other two $x_{\mathrm{R}}$, $x_{%
\mathrm{L}}$ is the saddle point. The transition boundary is then decided by
\begin{equation}
\varepsilon \left( x_{\mathrm{R}}\right) =\varepsilon \left( x_{\mathrm{L}%
}\right)
\end{equation}%
which leads us to
\begin{equation}
\left\vert g_{1c}\right\vert =g_{\mathrm{s}}[1+\frac{g_{\mathrm{t}}\epsilon
}{\widetilde{g}_{2}\Omega }]\sqrt{1-\widetilde{g}_{2}^{2}/g_{\mathrm{t}}^{2}}
\label{gc-low-w-limit}
\end{equation}%
in Eqs. \eqref{g1cSemiclassical} and \eqref{hzcSemiclassical}.

Note that, as mentioned for Fig. \ref{Fig-SemiClassicDiagrams} (d), in the
regime of negative $g_{2}$, the above boundary \eqref{gc-low-w-limit} is an
arc. Along this arc boundary the transition is of first order. At the ends
of the arc the transition becomes second order and boundary closes. As
revealed in Section \ref{Sect-E-SemiClasiscal}, the first-order transition
arises from the energy saddle point $x_{\mathrm{S}}$ which separates two
competing minima. Disappearing of the energy saddle will mean fading away of
the first-order transition. The critical point comes with a flattened
saddle. We show this saddle flattening in Fig. \ref{Fig-Saddle-flat}(a).
Here, the green dashed line illustrates the minimum-saddle-minimum of the
variational energy $\varepsilon $ at a point along the first-order boundary,
while the gray solid line shows the situation at the ends of the boundary
where a flattened bottom can be clearly seen. This critical point can be
figured out by vanishing of the first and second derivatives of the
variational energy
\begin{equation}
\frac{\partial \varepsilon \left( x\right) }{\partial x}=0,\quad \frac{%
\partial ^{2}\varepsilon \left( x\right) }{\partial x^{2}}=0.
\label{saddle-condition}
\end{equation}%
It should be mentioned \eqref{saddle-condition} is a necessary condition but
not a sufficient one. We give an example by the blue dot-dashed line in Fig. %
\ref{Fig-Saddle-flat}(a), where the middle point of the shoulder shape
fulfils \eqref{saddle-condition} but it is an inflection point instead of a
saddle point. Nevertheless, we can combine condition \eqref{saddle-condition}
and boundary \eqref{gc-low-w-limit} to extract the critical point,
\begin{equation}
g_{2}^{\mathrm{E}}\approx 3\left( \frac{\epsilon }{5\Omega }\right) ^{1/3}+%
\frac{226\epsilon }{75\Omega }-\frac{362011}{27000}\left( \frac{\epsilon }{%
5\Omega }\right) ^{5/3},  \label{g2E}
\end{equation}%
approximately for a weak bias and a non-linear interaction. Fig.\ref%
{Fig-Saddle-flat}(b) shows the above analytic $g_{2}^{\mathrm{E}}$ (dashed
line) in comparison with the numerical ones (dots). It is interesting to see
in the weak-bias regime $g_{2}^{\mathrm{E}}$ is in a fractional power law,
which means $g_{2}^{\mathrm{E}}$ increases quickly with a small strength of
the bias. A small bias could break and open much the ring of the round
boundary in Fig. \ref{Fig-SemiClassicDiagrams}.

\subsection{Analytic boundaries for the successive transitions at finite
frequencies}
\label{Sect-Analytic-Finite-w}

Based on the physical picture analyzed in Section \ref%
{Sect-tricritical-mechanism} we can obtain the phase boundaries in the
tricritical scenarios at finite frequencies. Unlike in the semiclassical
picture, now the left and right states can simultaneously get involved in a
ground state. We also decompose the wave function into right (R) and left
(L) states $\left\vert \Psi \right\rangle =c_{\mathrm{R}}\left\vert \psi _{%
\mathrm{R}}\right\rangle +c_{\mathrm{L}}\left\vert \psi _{\mathrm{L}%
}\right\rangle $\ upto a normalization factor. The right/left states are
respectively formed in the same-side tunneling $\Omega _{\alpha \beta }$ and
$\Omega _{\alpha \beta }$%
\begin{eqnarray}
\left\vert \psi _{\mathrm{L}}\right\rangle &=&\alpha ^{+}\varphi _{\alpha
}^{+}\left\vert \uparrow \right\rangle +\beta ^{-}\varphi _{\beta
}^{-}\left\vert \downarrow \right\rangle , \\
\left\vert \psi _{\mathrm{R}}\right\rangle &=&\alpha ^{-}\varphi _{\alpha
}^{-}\left\vert \downarrow \right\rangle +\beta ^{+}\varphi _{\beta
}^{+}\left\vert \uparrow \right\rangle ,
\end{eqnarray}%
where $\alpha ^{\pm },\beta ^{\pm }$ represent the weight of the wave packet
$\varphi _{j}^{\pm }$. The corresponding energy can be easily obtained as
\begin{eqnarray}
\varepsilon _{\mathrm{L}} &=&\frac{1}{2}\left[ (h_{\beta \beta
}^{-}+h_{\alpha \alpha }^{+})-\sqrt{(h_{\beta \beta }^{-}-h_{\alpha \alpha
}^{+})^{2}+S_{\alpha \overline{\beta }}^{2}\Omega ^{2}}\right] , \\
\varepsilon _{\mathrm{R}} &=&\frac{1}{2}\left[ (h_{\beta \beta
}^{+}+h_{\alpha \alpha }^{-})-\sqrt{(h_{\beta \beta }^{+}-h_{\alpha \alpha
}^{-})^{2}+S_{\beta \overline{\alpha }}^{2}\Omega ^{2}}\right] .
\end{eqnarray}%
Here we define $h_{ij}^{\pm }=\langle \varphi _{\mathrm{i}}^{\pm }|\left(
h^{\pm }-b_{0}-\varepsilon _{0}\right) |\varphi _{\mathrm{j}}^{\pm }\rangle $%
, where the irrelevant constants $b_{0}$ and $\varepsilon _{0}$ have been
substracted, and $S_{i\overline{j}}=\langle \varphi _{\mathrm{i}%
}^{+}|\varphi _{\mathrm{j}}^{-}\rangle $ is the wave-packet overlap. The
wave packet $\varphi _{\mathrm{j}}^{\pm }$ can be well approximated by the
displaced ground state of quantum harmonic oscillator, with the displacement
$\zeta _{i,\pm }x_{0,\pm }$ renormalized from the position of the potential
bottom $x_{0,\pm }$\cite{Ying2015}.\ Explicitly we have%
\[
h_{ii}^{\pm }=\frac{\omega }{2}\left\{ \varpi _{\pm }-\frac{[1-(1-\zeta
_{i,\pm })^{2}]g_{1}^{\prime 2}}{(1\pm \widetilde{g}_{2}^{\prime })}\right\}
\mp \epsilon ,
\]%
and $S_{\alpha \overline{\beta }}\approx S_{\beta \overline{\alpha }}\approx
1$ in gaining the maximum tunneling energy. The successive transitions occur
in weak bias and nonlinear interaction, in such situations we keep the
leading order
\begin{equation}
\varepsilon _{\mathrm{L}}-\varepsilon _{\mathrm{R}}=\overline{g}_{2}%
\overline{g}_{1}^{2}\zeta ^{2}\left[ \frac{\left( 1+\zeta /2\right) }{\sqrt{%
(\zeta ^{2}+\overline{g}_{1}^{-4})}}-1\right] \Omega -\frac{2\zeta \epsilon
}{\sqrt{(\zeta ^{2}+\overline{g}_{1}^{-4})}},  \label{eR-eL}
\end{equation}%
with $\zeta =(1-\overline{g}_{1}^{-4})^{1/2}$ being the displacement
renormalization from the conventional QRM \cite{Ying2015}.

Standing in a phase of the two-branch state, we can judge the onset of the
transitions to other states by an exponential decay of the state weight on
one side, $\delta _{c}=\left( c_{\mathrm{R}}/c_{\mathrm{L}}\right) ^{\pm
1}\sim e^{-1}$, where $\pm 1$ depends on which broken-branch state the
system is transiting to. Thus at the transition we can treat by a
perturbation from the left-right tunneling energy ($\Omega _{\alpha \alpha }$%
, $\Omega _{\beta \beta })$ as well as the single-particle left-right
overlap energy ($t_{\alpha \beta }^{+},t_{\beta \alpha }^{-})$
\begin{equation}
\delta _{c}=(\Omega _{\alpha \alpha }+\Omega _{\beta \beta }+t_{\alpha \beta
}^{+}+t_{\beta \alpha }^{-})/\left[ \eta _{\mathrm{LR}}\left( \varepsilon _{%
\mathrm{L}}-\varepsilon _{\mathrm{R}}\right) \right] ,  \label{deltaC}
\end{equation}%
where $\eta _{\mathrm{LR}}=\pm 1$ is decided by which side of state has a
lower energy. In the leading order, we have
\begin{eqnarray}
\Omega _{\alpha \alpha }+\Omega _{\beta \beta } &\approx &-\frac{\Omega }{2}%
S_{\alpha \overline{\alpha }}, \\
t_{\alpha \beta }^{+}+t_{\beta \alpha }^{-} &\approx &\alpha \beta \left[
\omega +\left( 1-\zeta \right) ^{2}\overline{g}_{1}^{2}\frac{\Omega }{2}%
\right] S_{\alpha \beta },
\end{eqnarray}%
where $\alpha =\sqrt{\left( 1+\zeta \right) /2}$ and $\beta =\sqrt{\left(
1-\zeta \right) /2}$ from the conventional QRM \cite{Ying2015} are the
leading contributions for $\alpha ^{\pm },\beta ^{\pm }$ which get involved
via $t_{ij}^{\pm }=w_{ij}h_{ij}^{\pm }$ , with $w_{ij}$ being the weight
product of $\alpha ^{\pm }$ and $\beta ^{\pm }.$ $S_{\alpha \beta }\approx
S_{\alpha \overline{\alpha }}\approx \exp [-\zeta ^{2}\overline{g}%
_{1}^{2}\Omega /(2\omega )]$ is approximate left-right wavepacket overlap.

Combining \eqref{eR-eL} and \eqref{deltaC}, we get analytic expressions for
boundaries II and III
\begin{equation}
\widetilde{g}_{2c}^{\mathrm{II,III}}=\pm \frac{(1-t)g_{\mathrm{t}}}{\delta
_{c}\zeta ^{3}\overline{g}_{1}^{2}}\exp [-\frac{\zeta ^{2}\overline{g}%
_{1}^{2}\Omega }{2\omega }]+\frac{4\epsilon }{\zeta ^{2}\overline{g}%
_{1}^{2}\Omega }g_{\mathrm{t}}.
\end{equation}%
Transition-IV is the shifting between pure left state and pure right state,
thus setting $\varepsilon _{\mathrm{L}}-\varepsilon _{\mathrm{R}}=0$ we find%
\begin{equation}
\widetilde{g}_{2c}^{\mathrm{IV}}=\frac{4\epsilon }{\zeta ^{2}\overline{g}%
_{1}^{2}\Omega }g_{\mathrm{t}}+O[\left( \frac{\epsilon }{\Omega }\right)
^{3}].
\end{equation}%
As we have seen from Figs. \ref{Fig-Frequency},\ref{Fig-tricritical-1},\ref%
{Fig-tricritical-2},\ref{Fig-Tricri-w-2},\ref{Fig-quadruple} in Section \ref%
{SectionAll-Finte-w}, these analytic boundaries work quite well in
comparison with the numerics.

\section{Conclusions and discussions}
\label{Sect-Conclusion}

By combining exact diagonalization and analytic methods in a semiclassical
picture and a full quantum-mechanical picture, we have presented a thorough
study on the ground state of the quantum Rabi model in the presence of the
bias and the nonlinear interaction. The model exhibits different patterns of
symmetry breaking, including the paramagnetic-like, antiferromagnetic like,
spontaneous symmetry breaking, paramagnetic-like plus first/second-order
transitions, antiferromagnetic-like plus first/second-order transitions.
These symmetry-breaking patterns bring a rich and colorful world of phase
diagrams. We have obtained the full phase diagrams and the analytic phase
boundaries, both in the low frequency limit and at finite frequencies. Five
different situations for the occurrence of tricriticality are unveiled,
respectively: (i) induced by the competition of the linear coupling and
nonlinear interaction in the presence of the bias, in the low frequency
limit. (ii) induced by raising the frequency in the respective presence of
the nonlinear interaction or the bias. (iii) induced by the competition of
linear coupling with the nonlinear interaction or the bias, under fixed
finite frequencies. (iv) induced by the interplay of linear coupling with
both the nonlinear interaction and the bias, under fixed finite frequencies.
(iv) induced by varying the frequency in the interplay of the nonlinear
interaction and the bias. The system could have four different quantum
phases, we revealed that all four phases can meet to form quadruple points.
The low-frequency-limit phase boundary of nonlinear interaction in the
absence of bias turns out to be a quadruple line. In comparison with the
semiclassical low-frequency limit, the finite frequencies lead to more phase
transitions. By analyzing the energy competitions and monitoring the
essential changes of quantum states in the transitions, we have clarified
the semiclassical and quantum-mechanical mechanisms underlying the
afore-mentioned phenomena. We see that the full quantum-mechanical effect
leads to much richer physics than the semiclassical picture, including
additional phase transitions, novel tricriticalities, and formation of
quadruple points as well as a fine structure of spontaneous symmetry
breaking.

Note that the model we consider can be implemented in the experimental
setups as in the superconducting circuit system\cite%
{Felicetti2018-mixed-TPP-SPP,Bertet-Nonlinear-Experim-Model-2005}. It is
convenient to cool the superconducting circuits down to the ground state. On
the other hand, the model parameters are controllable as the superconducting
systems are composed of LC circuits of which the frequency parameters are
quite tunable. It is worthwhile to give an estimation on the regime of
experimental parameters that is favorable for detections of the phenomena we
address in the present work, such as successive transitions and
tricricalities. The symmetry breaking patterns, second/first-order
transitions and tricriticality-(i) in the low frequency limit are
illustrated at frequencies of order $\omega =0.001\sim 0.01\Omega $, while
the nonlinear interaction $g_{2}$ has an order similar to $\omega $ and the
bias is in a range of order around $\epsilon =0\sim 10\omega .$ A typical
experimental strength for the tunneling strength $\Omega $ is of order $10$%
GHz \cite{Blais-2004-exp-parameters} in superconducting circuit systems,
although the order can reach $50$GHz in microwave cavities and even $350$THz
in optical cavities. For the superconducting systems we are more concerned,
the frequency $\omega =0.001\sim 0.01\Omega $ corresponds to the order $%
10\sim 100$MHz. The additional transitions and the novel tricricalities
occur at the finite frequencies of order around $\omega =0.1\Omega \sim
1\Omega $, while $\epsilon $ and $g_{2}$ are illustrated in a range of $%
10^{-5}\sim 10^{-1}g_{\mathrm{t}}$ where $g_{\mathrm{t}}$ is of the same
order of $\omega $. In LC circuits these parameters would correspond to $%
\omega =1\sim 10$GHz and $\epsilon ,g_{2}=10^{-2}\sim 10^{3}$MHz. These
parameter regimes would open a wide window accessible for the circuit
systems.

Our results would be relevant for the
growing interest in the nonlinear effect\cite{Ying-2018-arxiv,Felicetti2015-TwoPhotonProcess,Puebla2017-TwoPhotonProcess,Felicetti2018-mixed-TPP-SPP,Bertet-Nonlinear-Experim-Model-2005, Casanova2018npj,Puebla2019,CongLei2019,Xie2019,ZhengHang2017,PengJie2019,FelicettiPRL2020}
in the context of continuing enhancements of experimental light-matter
couplings\cite{Diaz2019RevModPhy,Wallraff2004,Gunter2009,Niemczyk2010,Peropadre2010,FornDiaz2017, Forn-Diaz2010,Scalari2012,Xiang2013,Yoshihara2017NatPhys,Yoshihara2017,Kockum2017}. Our analytic phase boundaries and physical analysis may provide some convenience and insights.
We speculate that the phenomena revealed here on
the classical-to-quantum transition and nonlinearity might also leave
imprints in the Bloch-Siegert effect\cite{Pietikainen2017} and dynamics\cite%
{Wolf2012,Hwang2015PRL,Crespi2012}, which we shall discuss in some other works.

\section*{Acknowledgements}

%\textbf{Acknowledgements}

This work was supported by the National Natural Science Foundation of China
(Grant No. 11974151).

\end{document}